\documentclass[twocolumn,showpacs,preprintnumbers,amsmath,amssymb]{revtex4-1}

\usepackage{graphicx}
\usepackage{dcolumn}
\usepackage{simplewick}
\usepackage{bm}
\usepackage{hyperref}
\usepackage[usenames,dvipsnames]{color}

\def\red{|\!|}
\def\bfd{\mathbf{D}}
\def\pt{\mathbf{T}^{(1)}}
\def\pto{\mathbf{T}^{(1)}_1}
\def\ptt{\mathbf{T}^{(1)}_2}
\def\ps{\mathbf{S}^{(1)}}
\def\pso{\mathbf{S}^{(1)}_1}
\def\pst{\mathbf{S}^{(1)}_2}
\def\helec{{\mathbf{H}}_{\rm elec}^{\rm NSD}}

\begin{document}

\title{Relativistic coupled-cluster theory of nuclear spin-dependent
       parity non-conservation }

\author{B. K. Mani}
\email{bkmani@prl.res.in}
\author{D. Angom}
\email{angom@prl.res.in}
\affiliation{Physical Research Laboratory,
             Navarangpura-380009, Gujarat,
             India}

\begin{abstract}
  We have developed a relativistic coupled-cluster theory to incorporate
nuclear spin-dependent interaction Hamiltonians perturbatively. In this
theory, the coupled-cluster operators in the electronic sector are defined
as tensor operators of rank one and we introduce suitable diagrammatic 
representations. For properties calculations, the electronic part is
first calculated and later coupled to the nuclear spin part. The 
method is ideal to calculate parity violating nuclear spin-dependent electric
dipole transition amplitudes, $E1_{\rm PNC}^{\rm NSD}$, of heavy atoms. To
validate the proposed method the $E1_{\rm PNC}^{\rm NSD}$ of the transition 
$6\; ^2S_{1/2} \rightarrow 7\; ^2S_{1/2}$ in 
$^{133}$Cs is calculated for selected MBPT diagrams and compared with the 
results from our theory. 
\end{abstract}



\maketitle


\section{Introduction}

 Experimental observation of nuclear spin dependent (NSD) parity 
non-conservation (PNC) is a signature of nuclear anapole moment (NAM) 
\cite{zeldovich-58}. It is the most dominant NSD-PNC effect among 
three possible sources. The other two are: combination of hyperfine and 
nuclear spin-independent PNC; and spin-dependent $Z$ exchange between electrons
and nucleus. To date, the most precise atomic PNC measurement, of atomic Cs, 
has provided the only experimental evidence of NAM \cite{wood-97}. One 
major hurdle to a clear observation of nuclear anapole moment is the large 
nuclear spin-independent (NSI) signal, which overwhelms the NSD signature. 
However, proposed experiments with single Ba$^+$ ion \cite{fortson-93} could 
probe PNC in the $s_{1/2}-d_{5/2}$ transition, where the NSI component is 
zero. This could then provide an unambiguous observation of NSD-PNC and 
NAM in particular. The ongoing experiments with atomic Ytterbium 
\cite{tsigutkin-09} is another possibility, the $6s^2\; ^1S_0-6s5d\; ^3D_2 $ 
transition, to observe NSD-PNC with minimal mixture from the NSI component. 
One crucial input, which is also the source of large uncertainty, to 
extract the value of NAM and  nuclear weak charge in NSI-PNC is the input 
from atomic theory calculations. In the case of isotope chain measurements, 
like in  Yb, the PNC observable is a ratio. Atomic theory contributions then 
cancel and parameters can be extracted without atomic theory calculations. 
However, results from atomic theory calculations are important in estimating 
the expected value of PNC transition amplitudes and extracting NAM does 
require results from atomic theory. For these reasons, it is important to 
employ reliable and accurate many-body theory in the atomic theory 
calculations.

  Recently, atomic theory calculations have investigated the NSD-PNC of 
Ba$^+$ and Ra$^+$ using the atomic many-body perturbation theory (MBPT)
\cite{sahoo-11}. Another recent work reports the theoretical estimate of the 
NSD-PNC observable of Yb \cite{dzuba-11a} and the calculations are based on 
the CI-MBPT method \cite{dzuba-96}. There is also an earlier work on NSD-PNC 
of Yb using the same method \cite{porsev-00}. Very recently, the CI-MBPT method
is used to  to calculate the NSD-PNC observables of Ba$^+$, Yb$^+$ and Ra$^+$
\cite{dzuba-11b}. Besides the methods mentioned, earlier works on Ba$^+$ 
\cite{geetha-98} and Yb \cite{angom-99} used configuration interaction (CI). 
To date, coupled-cluster theory (CCT) considered as one of the most reliable 
and accurate many-body theory has not been used in the NSD-PNC calculations.
The difficulty of developing a suitable theory stems from the complications
of dealing with the nuclear spin dependent interaction. In a previous paper of
ours we reported the development of a relativistic coupled-cluster (RCC) 
theory to incorporate nuclear spin-dependent interaction in a consistent 
scheme \cite{mani-11b}. In this work we provide
elaborate details of the theory and touch upon related subtle issues. 

    The coupled-cluster (CC) theory\cite{coester-58,coester-60} 
is one of the most reliable many-body theory to incorporate electron 
correlation in atomic calculations. In atomic physics, the relativistic 
coupled-cluster (RCC) theory has been used 
extensively in atomic properties calculations, for example, hyperfine 
structure constants \cite{pal-07,sahoo-09}, electric dipole moments 
\cite{nataraj-08,latha-09}, and electromagnetic 
transition properties \cite{thierfelder-09,sahoo-09a}. In atomic PNC 
calculations too, RCC is the preferred theory and several groups have
used it to calculate NSI-PNC of atoms \cite{wansbeek-08,pal-09,porsev-10}. 
However, the calculations in Ref. \cite{wansbeek-08} are entirely based on
RCC with a variation we refer to as perturbed RCC (PRCC), where as 
the calculations in Ref. \cite{pal-09,porsev-10} are based on sum over states 
with CC wave functions. Naturally, the former incorporates electron correlation
more precisely than the later approach. 

 In this work we provide a detailed explanation on the approach
we have adopted to incorporate nuclear spin-dependent interaction Hamiltonian
as a perturbation in RCC theory. In Section. \ref{review_rcc}, we give a very
short account of RCC theory to serve as a quick reference. We then give
the details of our formulation to represent a NSD interaction perturbed 
CC operator in Section. \ref{pert_cc_fn}. PRCC equations of closed-shell and 
one-valence  systems is derived in this section. To simplify angular factor
calculations, the calculations in the electronic sector is separated out.
Based on this the calculation of NSD-PNC observable, in the electronic sector, 
is described in Section. \ref{e1pnc_elec}. Next section, 
Section. \ref{couple_nsd}, describes the coupling of results from the 
electronic sector with the nuclear spin. This completes the theoretical
development and then, the correctness of theory is verified in 
Section. \ref{validation_prcc}. For which, we use MBPT results and calculate
the NSD-PNC of the $6\;^2S_{1/2}\rightarrow 7\;^2S_{1/2}$ transition in Cs. 
We then conclude and the angular factors of all the diagrams in the 
linearized PRCC are given in the Appendix.


\section{Brief review of RCC}
\label{review_rcc}

   For $N$-electron atoms or ions the Dirac-Coulomb Hamiltonian, 
appropriate to account for relativistic effects, is  
\begin{equation}
  H^{\rm DC}=\sum_{i=1}^N\left [c\bm{\alpha}_i\cdot \mathbf{p}_i+
             (\beta_i-1)c^2 - V_N(r_i)\right ] +\sum_{i<j}\frac{1}{r_{ij}},
  \label{dchamil}
\end{equation}
where $\bm{\alpha}_i$ and $\beta$ are the Dirac matrices, $\mathbf{p}$ is the
linear momentum, $V_N(r)$ is the nuclear Coulomb potential and last term
is the electron-electron Coulomb interactions. For one-valence systems it 
satisfies the eigen value equation
\begin{equation}
  H^{\rm DC}|\Psi_v\rangle = E_v|\Psi_v\rangle,
  \label{dcpsi}
\end{equation}
where $|\Psi_v\rangle$ and $E_v$ are the atomic state and energy respectively. 
In the CC method, the atomic state is expressed in terms of $T$  and $S$, the 
closed-shell and one-valence cluster operators respectively, as
\begin{equation}
  |\Psi_v\rangle = e^{T^{(0)}} \left [  1 + S^{(0)} \right ] |\Phi_v\rangle,
  \label{cceqn_1v}
\end{equation}
where $|\Phi_v\rangle$ is the one-valence Dirac-Fock reference state. It is 
obtained by adding an electron to the closed-shell reference state,
$|\Phi_v \rangle = a^\dagger_v|\Phi_0\rangle$. In the coupled-cluster singles 
doubles (CCSD) approximation $T^{(0)} = T^{(0)}_1 + T^{(0)}_2$ and are the 
solutions of the nonlinear coupled equations
\begin{subequations}
\label{t_eqn}
\begin{eqnarray}
  \langle\Phi^p_a|\bar H_{\rm N}|\Phi_0\rangle = 0, 
     \label{t1_eqn}                        \\
  \langle\Phi^{pq}_{ab}|\bar H_{\rm N}|\Phi_0\rangle = 0,
     \label{t2_eqn} 
\end{eqnarray}
\end{subequations}
where $\bar H_{\rm N}=e^{-T^{(0)}}H_{\rm N}e^{T^{(0)}} $ is the 
similarity transformed Hamiltonian and  the normal order Hamiltonian
$H_{\rm N} = H -\langle\Phi_0|H|\Phi_0\rangle$.  The states $|\Phi^p_a\rangle$ 
and $|\Phi^{pq}_{ab}\rangle$ are the singly and doubly excited determinants,
respectively and $abc\ldots$ ($pqr\ldots$) denote occupied (virtual) orbitals. 
The details of the derivation are given in Ref. \cite{mani-09}. 
Like in $T^{(0)}$, the one-valence cluster operator $S^{(0)}$ in the CCSD 
approximation is $S^{(0)} = S^{(0)}_1 + S^{(0)}_2$. And these are the 
solutions of
\begin{subequations}
\label{cc_sin_dou}
\begin{eqnarray}
  \langle \Phi_v^p|\bar H_N \! +\! \{\contraction[0.5ex]
  {\bar}{H}{_N}{S} \bar H_N S^{(0)}\} |\Phi_v\rangle
  &=&E_v^{\rm att}\langle\Phi_v^p|S^{(0)}_1|\Phi_v\rangle ,
  \label{ccsingles}     \\
  \langle \Phi_{va}^{pq}|\bar H_N +\{\contraction[0.5ex]
  {\bar}{H}{_N}{S}\bar H_N S^{(0)}\} |\Phi_v\rangle
  &=& E_v^{\rm att}\langle\Phi_{va}^{pq}|S^{(0)}_2|\Phi_v\rangle,
  \label{ccdoubles}
\end{eqnarray}
\end{subequations}
where $E_v^{\rm att} = E_v - E_0,$ is the attachment energy of the
valence electron. In our previous work \cite{mani-10}, we provide details
of the derivation.


\section{Perturbed CC wave function}
\label{pert_cc_fn}

   Time independent perturbation theory is the standard procedure to 
incorporate external perturbations or additional interactions in atomic 
many-body calculations. However, a basic requirement of perturbative 
calculations is a complete set of intermediate atomic states, which 
is non-trivial to generate. The perturbed CC method 
\cite{latha-08,mani-09,sahoo-08}, on the other hand, 
implicitly accounts for all the possible intermediate states. In this work,
we consider the perturbation as the nuclear spin-dependent parity 
non-conserving (PNC) interaction
\begin{equation}
   H_{\rm PNC}^{\rm NSD}=\frac{G_{\rm F}\mu'_W}{\sqrt{2}}\sum_i
   \bm{\alpha}_i\cdot \mathbf{I}\rho_{\rm{N}}(r),
  \label{hpncnsd2}
\end{equation}
where $\mu'_W$ is the weak nuclear moment of the nucleus and
$\rho_{\rm N}(r)$ is the nuclear density. The weak nuclear moment
is expressed in terms of the neutron and proton numbers
$\mu'_W = 2(Z C_{1p} + N C_{1n})$, where $C_{1p}$ and $C_{1n}$ are
respectively the vector electron and the axial vector nucleon coupling
coefficients. There are two complications arising from the nuclear spin 
operator $I$ in $H_{\rm PNC}^{\rm NSD}$. First, the cluster operators in the 
electron space are rank one operators, and second, the atomic states in the 
one-valence sector are eigenstates of total angular momentum
$\mathbf{F} = \mathbf{I} + \mathbf{J}$. Both of these are relatively simple
to incorporate at lower order MBPT calculations, however, implementing these
in non-perturbative theory like RCC is nontrivial. Accordingly, the method we 
have developed and implemented in the current work are very different from 
our previous works \cite{latha-08,mani-09}. The other simplifying feature of 
the previous works is, the methods developed were for closed-shell systems 
where the total electronic angular momentum $\mathbf{J} = 0$.

  With the PNC interaction, the total atomic Hamiltonian is 
\begin{equation}
    H_{\rm A} = H^{\rm DC} + \lambda H_{\rm PNC}^{\rm NSD},
    \label{total_H}
\end{equation}
where $\lambda$ is the perturbation parameter. As $H_{\rm PNC}^{\rm NSD}$ 
mixes atomic states of opposite parities, the eigenvalue equation is modified to
\begin{equation}
  H_{\rm A}|\widetilde{\Psi}_v\rangle = E_v|\widetilde{\Psi}_v\rangle.
  \label{pnc_eigen}
\end{equation}
Note that the first order energy correction 
$E^1 = \langle\Psi_0|H_{\rm PNC}^{\rm NSD}|\Psi_0\rangle = 0$
as $H_{\rm PNC}^{\rm NSD}$ is an odd parity interaction Hamiltonian. This is 
taken into account while writing the above eigenvalue equation. Here 
$ |\widetilde{\Psi}_v\rangle$ are the mixed parity atomic states and to first 
order in perturbation
\begin{equation}
   |\widetilde{\Psi}_v \rangle = |\Psi_v \rangle +
   \sum_I |\Psi_I\rangle \frac{\langle\Psi_I|H_{\rm PNC}^{\rm NSD}|
           \Psi_v\rangle} {E_v - E_I}.
  \label{psi_pert}
\end{equation}
If $|\Psi_v\rangle$ and $|\Psi_w\rangle$ are atomic states of same parity, then
the $H_{\rm PNC}^{\rm NSD}$ induced electric dipole ($E1$ ) transition 
amplitude is 
\begin{equation}
  E1_{\rm PNC}^{\rm NSD}  =  \langle \widetilde{\Psi}_w \red \mathbf{D} \red
                  \widetilde{\Psi}_v \rangle, 
  \label{e1pnc_general}
\end{equation}
where $\mathbf{D}$ is the dipole operator. The perturbation expression in 
Eq. (\ref{psi_pert}) require a complete set of intermediate atomic states 
$|\Psi_I\rangle$. This, as mentioned at the beginning of the section, is 
non-trivial to obtain in atomic many-body calculations. Summation
over intermediate states is circumvented when $E1_{\rm PNC}^{\rm NSD}$ is 
calculated with  CC atomic states. For this define a new set of cluster 
operators $\pt$, which unlike $T^{(0)}$ connects the reference state to 
opposite parity states. This is the result of incorporating one order of 
$H_{\rm PNC}^{\rm NSD}$ and for this reason we refer to $\pt $ as 
the perturbed cluster operators. Although hyperfine states are natural to
$H_{\rm PNC}^{\rm NSD}$,  cluster operator $\pt $ is defined to 
operate only in the electronic space and is a rank one operator. So, the 
mixed parity atomic state in RCC is 
\begin{equation}
 |\widetilde{\Psi}_0 \rangle  =  e^{T^{(0)}+ \lambda\pt\cdot\mathbf{I}}
                                |\Phi_0\rangle. 
 \label{prcc_cl}
\end{equation}
The scalar product with the nuclear spin $\mathbf{I}$ in the 
exponent restores $\pt $ to the correct form of the wave operator. 
At this point it is convenient to separate out the electronic part of the 
interaction Hamiltonian 
\begin{equation}
   \helec = \frac{G_{\rm F}\mu'_W}{\sqrt{2}}\sum_i
            \vec \alpha_i\rho_{\rm{N}}(r),
\end{equation}
which operates only in the electronic space,  so that
\begin{equation}
   H_{\rm PNC}^{\rm NSD}=  \helec\cdot \mathbf{I}.
\end{equation}
The remaining part of this section describe how to arrive at a consistent 
representation of $\pt$ and extension of the method to one-valence systems, 
where $\ps$ are the perturbed cluster operators.


\subsection{MBPT wavefunctions}

 To define the multipole and parity selection rules of $T^{(1)}$, we examine 
the second order MBPT wave function with $H_{\rm PNC}^{\rm NSD}$  as one of the 
perturbations. From the generalized Bloch equation \cite{lindgren-85}, the 
total wave operator is \cite{sahoo-11}
\begin{equation}
 \Omega_v = \sum_{n=1}^{\infty}\Omega_{v, 0}^{(n)}
            + \sum_{n=1}^{\infty}\Omega_{v, 1}^{(n)}. 
\end{equation}
Here $\Omega_{v, 0}^{(n)}$ has $n$ orders of residual Coulomb interaction, 
where as $\Omega_{v, 1}^{(n)}$ has one order of $H_{\rm PNC}^{\rm NSD}$ and 
$n$ orders of residual Coulomb interaction. Following which the mixed parity 
state 
\begin{equation}
   |\widetilde{\Psi}_v \rangle = |\Psi_v \rangle + |\bar \Psi_v \rangle
   = \sum_{n=1}^{\infty}\Omega_{v, 0}^{(n)} |\Phi_v\rangle
   + \sum_{n=1}^{\infty}\Omega_{v, 1}^{(n)} |\Phi_v\rangle , 
\end{equation}
where $|\bar \Psi_v \rangle $ is the opposite parity component arising from 
the $H_{\rm PNC}^{\rm NSD}$. The first order wave function is then,
\begin{equation}
   |\widetilde{\Psi}_v^1 \rangle = |\Psi_v^1 \rangle + |\bar\Psi_v^1 \rangle
      = \left[ 1 + \Omega^{(1)}_{v,0} \right] |\Phi_v \rangle 
      + \Omega^{(1)}_{v, 1} |\Phi_v \rangle.
 \label{pert_state_mbpt}
\end{equation}
Although, notation wise $\Omega_{v, 1}^{(1)}$ seem first order, it is 
second order in perturbation: one order each in residual Coulomb interaction
and $H_{\rm PNC}^{\rm NSD}$. For model space consisting of same parity states
\begin{equation}
   \left [ \Omega_{v,1}^{(1)}, H_0 \right ] P = QH_{\rm PNC}^{\rm NSD}
      \Omega_{v,0} P + QV_{\rm res}\Omega_{v,1}^{(0)} P.
  \label{omega_11}
\end{equation}
Here, and  $P$ and $Q$ are the projection operators of the model and 
complementary spaces, respectively, $H_0$ is the Dirac-Fock Hamiltonian, 
$V_{\rm res}$ is the residual Coulomb interaction and the zeroth order mixed 
parity wave operator
\begin{equation}
  \Omega_{v, 1}^{(0)}P = Q\frac{1}{E_v^0 - H_0}QH_{\rm PNC}^{\rm NSD}P .
 \label{omega_v1} 
\end{equation}
Using $\Omega_{v,1}^{(1)}$ and $\Omega_{v, 0}^{(1)} $, we may compute 
$E1_{\rm PNC}$ to third order in perturbation, the details of which are
discussed in Ref. \cite{sahoo-11}.


\subsection{Perturbed cluster operator representation}
\label{pert_cl_rep}

The expression of $\Omega_{v, 1}^{(1)} $ and associated selection rules are 
what we need to arrive at a consistent description of $T^{(1)} $.  
%
%
\begin{figure}[h]
\begin{center}
  \includegraphics[width = 7.6cm]{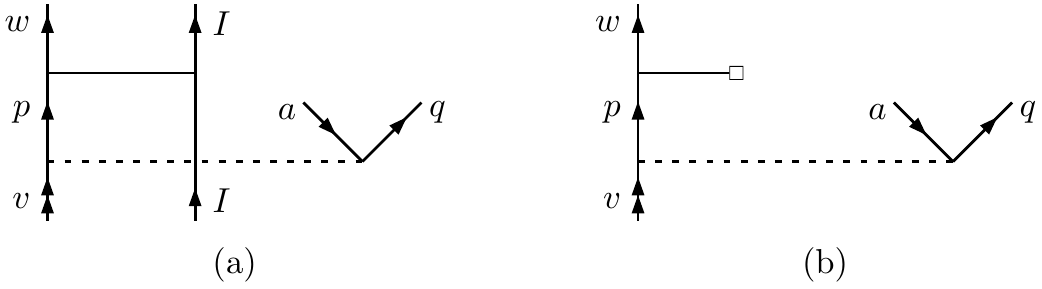}
  \caption{Second order MBPT diagram where interaction Hamiltonian 
           $H_{\rm PNC}^{\rm NSD}$ operates on the particle state $p$.
           (a) Diagrammatic representation of $H_{\rm PNC}^{\rm NSD}$ in
           the electronic and nuclear spin space. (b) Diagrammatic
           representation within the electronic space only where the 
           PNC interaction Hamiltonian is $H_{\rm elec}^{\rm NSD}$. Line 
           terminated with square represents the PNC interaction.}
  \label{pcc_mbpt}
\end{center}
\end{figure}
In Fig. \ref{pcc_mbpt}(a) we show an MBPT wave function diagram of 
$\Omega_{v, 1}^{(1)}$ arising from the second term in Eq. (\ref{omega_v1}). 
The algebraic expression of the diagram is 
\begin{equation}
  \frac{G_{\rm F}\mu'_W}{\sqrt{2}}
  \sum_p |wqI\rangle \frac{\langle w I |\bm{\alpha} \cdot \mathbf{I} 
    \rho_{\rm{N}}(r)| p I \rangle \langle pq I |V_{\rm res}|va I \rangle}
       { (\epsilon_{w} - \epsilon_{p}) 
       (\epsilon_{v} + \epsilon_{a} - \epsilon_{p} - \epsilon_{q})},
\end{equation}
where $\epsilon_i$ are the orbital energies. The states $|\cdots I \rangle$ 
represent uncoupled atomic states comprising of electronic and nuclear states.

%
%
\begin{figure}[h]
\begin{center}
  \includegraphics[width = 8.0cm]{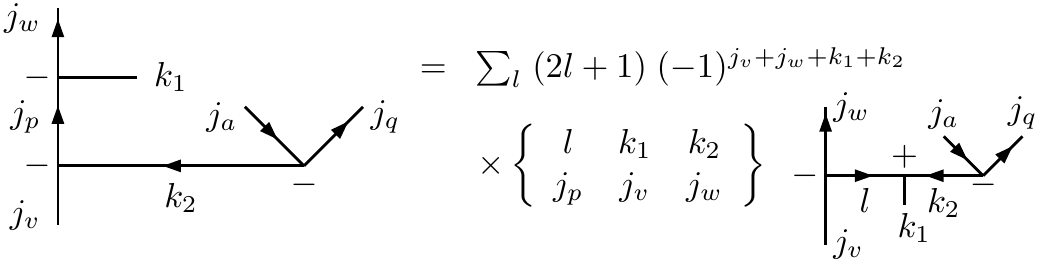}
  \caption{Angular reduction of the second order MBPT diagram. The remnant 
           angular momentum diagram on the right is representative of the 
           multipole structure of PRCC double excitation cluster operators
           $T_2^{(1)} $ and $S_2^{(1)}$.}
  \label{pcc_mbpt1}
\end{center}
\end{figure}

\begin{figure}
\begin{center}
  \includegraphics[width = 7.0 cm]{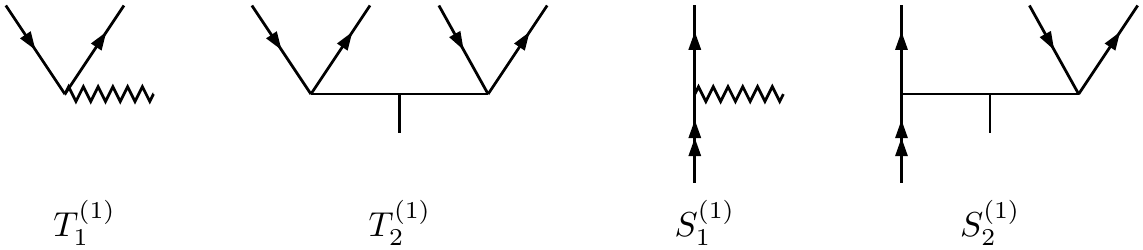}
  \caption{Diagrammatic representation of the single and double excitation 
           perturbed cluster operators in closed shell and one-valence 
           sectors. The extra line in the $T_2^{(1)} $ and $S_2^{(1)}$ is to
           indicate the multipole structure of the operators.}
  \label{pert_cc_op}
\end{center}
\end{figure}

On closer examination, the component of $H_{\rm PNC}^{\rm NSD}$ which 
operates in the nuclear subspace is diagonal. So, for further calculations, we 
separate out the matrix elements in the electronic subspace and combine 
with the nuclear part at the end of the calculations. For the MBPT diagram 
considered earlier Fig. \ref{pcc_mbpt}(a), the electronic part is as shown in 
Fig. \ref{pcc_mbpt}(b). The corresponding algebraic expression is
\begin{equation}
  \frac{G_{\rm F}\mu'_W}{\sqrt{2}}
  \sum_{p} \frac{\langle w |\bm{\alpha} \rho_{\rm{N}}(r)| p \rangle 
  \langle pq |V|va \rangle} { (\epsilon_{w} - \epsilon_{p}) 
  (\epsilon_{v} + \epsilon_{a} - \epsilon_{p} - \epsilon_{q})}.
 \label{mbpt_2body}
\end{equation}
Details of the angular reduction when electronic state is coupled with the 
nuclear state are  discussed in the later sections of the paper. Like wise, 
the perturbed cluster operators $\pt$, as defined earlier operate only in 
the electronic. For this consider the angular part of the matrix elements in  
Eq. (\ref{mbpt_2body}), the diagrammatic representation is shown in 
Fig. \ref{pcc_mbpt1}. The angular diagrams are based on the conventions used
in Lindgren and Morrison \cite{lindgren-85} and the same is followed in the
remaining of the paper while referring to angular momentum diagrams. In the 
figure, the angular diagram on the right hand side indicates the multipole 
structure of the $T^{(1)}_2$ and like the electron-electron Coulomb 
interaction
\begin{equation}
   \ptt = \sum_{abpq}\sum_{l, k_2} \tau_{ab}^{pq}(l, k_2) 
       \{\mathbf{C}_l(\hat{r}_1)\mathbf{C}_{k_2}(\hat{r}_2)  \}^1, 
\end{equation}
where $\mathbf{C}_i $ are c-tensor operators and $\{\cdots \}^1 $ indicates
the two c-tensor operators couple to a rank one tensor operator. Following
general rules of coupling tensor operators, the rank of the tensor operators 
must satisfy the triangular conditions
$|j_w - j_v| \leqslant l \leqslant (j_w + j_v)$, 
$|j_a - j_q| \leqslant k_2 \leqslant (j_a + j_q)$  and 
$|l - k_2| \leqslant 1 \leqslant (l + k_2)$. Effectively, $\pt$ is a 
rank one operator in the electronic subspace and  forms a scalar operator 
after coupling with $\mathbf{I}$. The other important difference from the 
$T^{(0)} $ is the parity selection rule at the vertices. The combined parities 
at the vertices are opposite 
$(-1)^{l_w + l_v} = - (-1)^{l_a + l_q}$ . After a similar analysis, the 
singles operator is
\begin{equation}
  \pto = \sum_{ap}\tau_a^p \mathbf{C}_1(\hat{r}).
\end{equation}
Based on the multipole structures, the perturbed cluster operators are 
diagrammatically represented as shown in Fig. \ref{pert_cc_op}. For the 
doubles $\ptt $, to indicate the multipole structure,  an additional 
line is added to the interaction line.


\subsection{Closed-shell systems}

\begin{figure}
\begin{center}
  \includegraphics[width = 7.0cm]{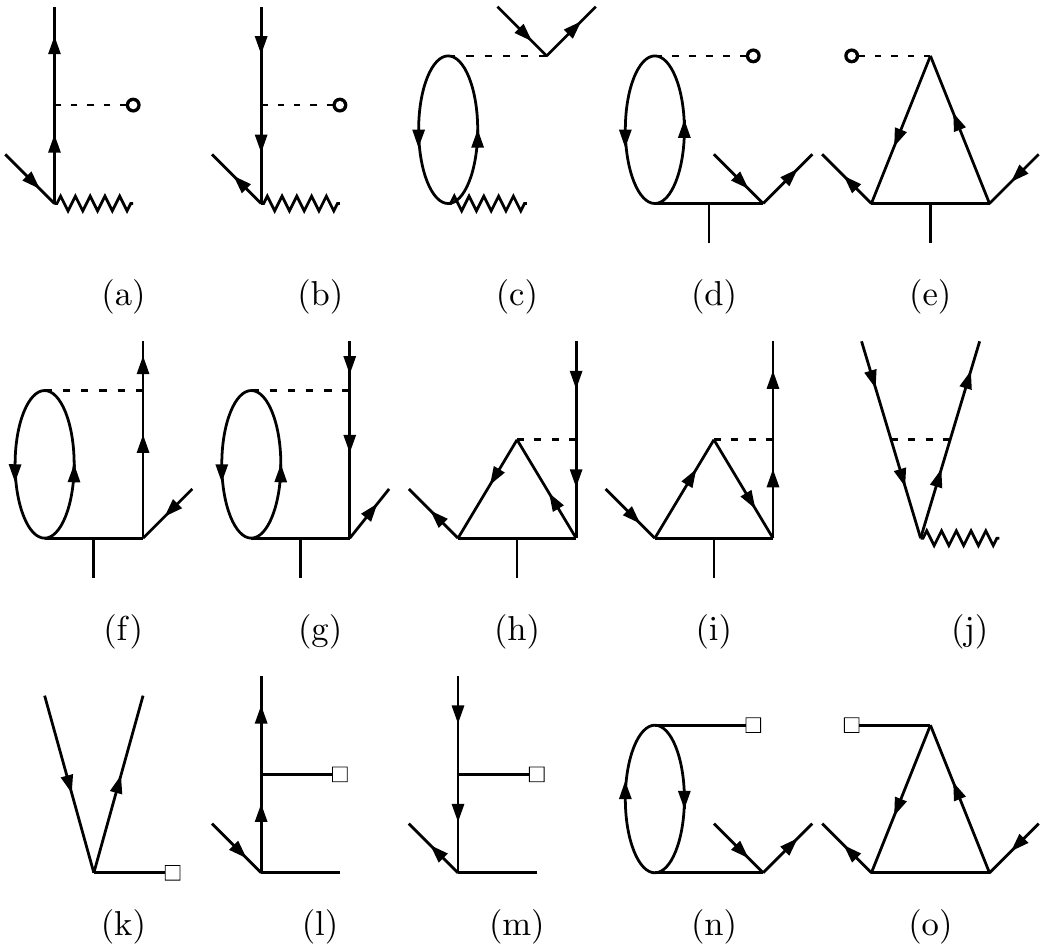}
  \caption{Diagrams which contribute to the linearized perturbed 
           coupled-cluster equations of the singly excited cluster operator
           $T_1^{(1)}$.}
  \label{psingles}
\end{center}
\end{figure}

\begin{figure}
\begin{center}
  \includegraphics[width = 7.0cm]{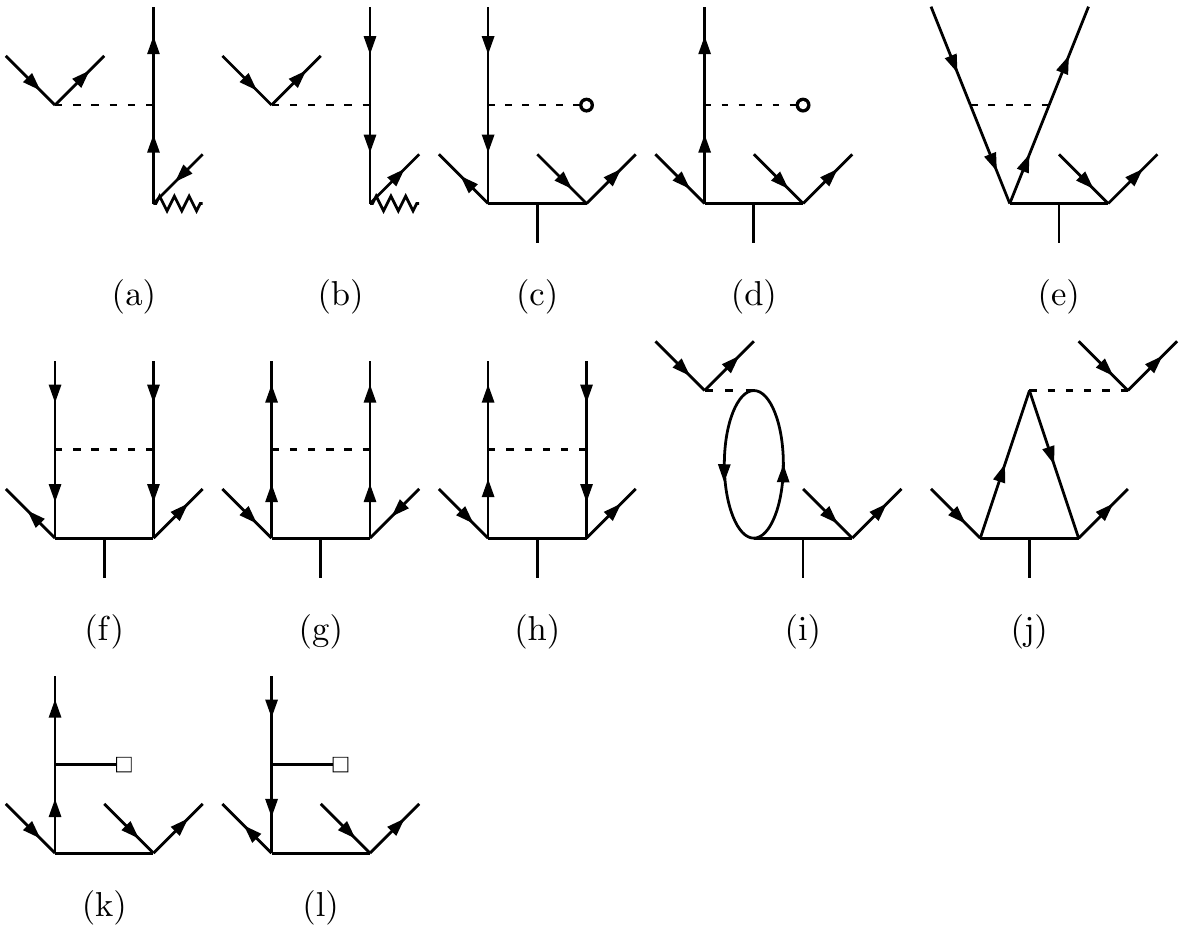}
  \caption{Diagrams which contribute to the closed-shell linearized perturbed 
           coupled-cluster equations of the doubly excited cluster operator
           $T_2^{(1)}$.}
  \label{pdoubles}
\end{center}
\end{figure}

The mixed parity states, eigenstates of $H_{\rm A}$, in the perturbed 
relativistic coupled-cluster (PRCC) is as given in Eq. (\ref{prcc_cl}).
However, as $H_{\rm PNC}^{\rm NSD}$ is considered to first order only,
it is sufficient to consider up to linear terms in the expansion of 
$e^{\lambda\pt\cdot\mathbf{I}}$. The PRCC atomic state is then
\begin{equation}
  |\widetilde \Psi_0 \rangle = e^{T^{(0)}} \left [  1 
    + \lambda \pt\cdot\mathbf{I} \right ] |\Phi_0\rangle.
  \label{pccwave2}
\end{equation}
The cluster operator $\pt$, as described earlier, incorporates $\helec$ to 
first order and residual Coulomb interaction to all order. 
%
%
%
%
\vspace{0.2cm}
The closed-shell equivalent of Eq. (\ref{pnc_eigen}), the mixed parity
eigenvalue equation, is 
\begin{equation}
  \left ( H^{\rm DC} + \lambda \helec\cdot\mathbf{I}\right ) 
       |\widetilde{\Psi}_0\rangle = E_0|\widetilde{\Psi}_0\rangle.
  \label{pnc_grnd}
\end{equation}
From the definition of the coupled-cluster mixed parity state in 
Eq. (\ref{pccwave2}), the eigen value equation in terms of the coupled-cluster 
wavefunction is  
\begin{eqnarray}
  \left( H^{\rm DC} + \lambda \helec \cdot\mathbf{I}\right) e^{T^{(0)}}
  \left[ 1 + \lambda \pt\cdot\mathbf{I}  \right] |\Phi_0 \rangle 
                              \nonumber \\
   = E_0 e^{T^{(0)}}\left[ 1 + \lambda \pt\cdot\mathbf{I} \right]|\Phi_0\rangle.
\end{eqnarray}
In terms of $H_N$, defined earlier, the eigenvalue equation is simplified to
\begin{eqnarray}
  \left( H_{\rm N} + \lambda \helec \right) e^{T^{(0)}}
  \left[ 1 + \lambda \pt\cdot\mathbf{I} \right] |\Phi_0 \rangle
                              \nonumber \\
  = \Delta E_0 e^{T^{(0)}} \left[ 1 + \lambda \pt\cdot\mathbf{I} \right] 
     \Phi_0 \rangle,
\end{eqnarray}
where $\Delta E_0 = E_0 - \langle\Phi_0|H|\Phi_0\rangle$ is the closed-shell
correlation energy. Retaining terms which are first-order in $\lambda$,
we get
\begin{equation}
  \left[ H_{\rm N} \pt \cdot\mathbf{I} + \helec \cdot\mathbf{I}\right] 
   e^{T^{(0)}} |\Phi_0 \rangle = \Delta E_0 \pt\cdot\mathbf{I} e^{T^{(0)}} 
   |\Phi_0 \rangle.
\end{equation}
From here on, for simplicity, we drop the nuclear spin operator $\mathbf{I}$ 
and work only in the electronic space. At a later stage of the calculations
the electronic part is coupled with the nuclear spin. Operating with 
$e^{-T^{(0)}}$ and projecting on singly and doubly excited
states $\langle\Phi^p_a|$ and $\langle\Phi^{pq}_{ab}|$, respectively,
we get the CC equations for singles and doubles perturbed cluster
amplitudes as
\begin{subequations}
\label{pcceq}
\begin{eqnarray}
  \langle \Phi^p_a |\{ \contraction{}{H}{_{\rm N}}{T}
     \bar{\mathbf{H}}_{\rm N}\mathbf{T}^{(1)} \} |\Phi_0\rangle =
  -\langle \Phi^p_a | \bar {\mathbf{H}}_{\rm elec}^{\rm NSD} |\Phi_0 \rangle,
  \label{pcceq1}                         \\
  \langle \Phi^{pq}_{ab} | \{\contraction{}{H}{_{\rm N}}{T}
  \bar{\mathbf{H}}_{\rm N}\mathbf{T}^{(1)} \} |\Phi_0 \rangle =
  -\langle \Phi^{pq}_{ab} | \bar {\mathbf{H}}_{\rm elec}^{\rm NSD} |\Phi_0 
   \rangle
 \label{pcceq2}
\end{eqnarray}
\end{subequations}
Where, $\bar{\mathbf{H}}_{\rm elec}^{\rm NSD}=e^{-T^{(0)}}\helec e^{T^{(0)}}$, 
is the similarity transformed PNC interaction Hamiltonian in the electronic 
subspace. The perturbed cluster operators are solutions of these coupled 
linear equations. However, the equations are nonlinear in the unperturbed 
cluster operators $T^{(0)}$. The advantage of separating the cluster operators 
into two categories $T^{(0)} $ and $\pt$ is, the two sets of equations can be 
solved sequentially. As the $ T^{(0)}$ has no dependence on $\pt$, 
the $ T^{(0)}$ equations are solved first and then the $\pt$  equations
are solved. An approximate form of Eq. (\ref{pcceq1}) and (\ref{pcceq2}),
but which contains all the important many-body effects, are the linearized
cluster equations. This is obtained by considering
\begin{subequations}
\label{lpcc_term}
\begin{eqnarray}
   \contraction{}{H}{_{\rm N}}{T}\bar{H}_{\rm N}\mathbf{T}^{(1)} & \approx & 
        \contraction{}{H}{_{\rm N}}{T}{H}_{\rm N}\mathbf{T}^{(1)},  \\
   \bar {\mathbf{H}}_{\rm elec}^{\rm NSD} & \approx &  
      \mathbf{H}_{\rm elec}^{\rm NSD} + 
   \contraction{}{H}{_{\rm elec}^{\rm NSD}}{T}
       {\mathbf{H}}_{\rm elec}^{\rm NSD}T^{(0)}.
\end{eqnarray}
\end{subequations}
The diagrams in the singles and doubles equation arising from 
$\contraction{}{H}{_{\rm N}}{T}{H}_{\rm N}\mathbf{T}^{(1)}$  are shown in 
Fig. \ref{psingles} and \ref{pdoubles}, respectively.

%
%
%
%
\vspace{0.2cm}
\vspace{0.2cm}
The other form of Eq. (\ref{lpcc_term}) is to write the equation in
terms of specific cluster amplitudes. The singles equation is then, 
\begin{eqnarray}
  (\epsilon_a - &\epsilon_p&) \bm{\tau}^p_a = \bm{h}_{pa} 
      + \sum_q \bm{h}_{pq} t^q_a  -\sum_b \bm{h}_{ba} t^p_b 
      + \sum_{bq}\bm{h}_{bq}\widetilde t^{qp}_{ba} 
                  \nonumber \\
  &&  + \sum_{bq}{\widetilde g}_{bpqa} \bm{\tau}^q_b +
  \sum_{bqr}{\widetilde g}_{bpqr} \bm{\tau}^{qr}_{ba}  
  -\sum_{bcq} g_{bcqa} \bm{\widetilde\tau}^{qp}_{bc},
  \label{lin_s}        
\end{eqnarray}
where $g_{ijkl}$ and $\bm{h}_{ij}$, are the matrix elements  
$\langle ij|V|kl\rangle$ and 
$\langle i|\mathbf{H}_{\rm elec}^{\rm NSD}|j\rangle$, respectively, and
$t^i_j$ and $t^{ij}_{kl}$ are the unperturbed single and double excitation
RCC amplitudes, respectively. For compact notation, we have defined 
 ${\widetilde g}_{ijkl} = g_{ijkl} - g_{ijlk}$.
Similarly, ${\widetilde t}^{ij}_{kl}$ and ${\widetilde {\bm{\tau}}}^{ij}_{kl}$
are the antisymmetized unperturbed and perturbed CC amplitudes respectively. 
The equation of the double excitation perturbed cluster amplitudes is
\begin{eqnarray}
  (\epsilon_a + &\epsilon_b &- \epsilon_p - \epsilon_q) \bm{\tau}^{pq}_{ab} =
     \left ( \sum_r\bm{h}_{pr} t^{rq}_{ab}-\sum_c\bm{h}_{ca}t^{pq}_{cb} 
           \right .        \nonumber \\
     && +\sum_{r} g_{pqrb} \bm{\tau}^r_a  -\sum_{c} g_{cqab} \bm{\tau}^p_c 
  +\sum_{rc} g_{pcar} \bm{\widetilde\tau}^{rq}_{cb} 
                  \nonumber \\
  && \left .  -\sum_{rc} g_{pcrb} \bm{\tau}^{rq}_{ac}  
      -\sum_{rc} g_{cpar} \bm{\tau}^{rq}_{cb}  \right ) + 
     \left ( \begin{array}{c}
              p\leftrightarrow q \\
              a\leftrightarrow b 
             \end{array} \right ) 
                  \nonumber \\
  &&  + 
  \sum_{rs} g_{pqrs} \bm{\tau}^{rs}_{ab}
      +\sum_{cd} g_{cdab} \bm{\tau}^{pq}_{cd}.
  \label{lin_d}                         
\end{eqnarray}
where $\bigl( \begin{smallmatrix}p\leftrightarrow q \\ a\leftrightarrow b
\end{smallmatrix} \bigr )$ represents terms similar to those within 
parenthesis but with the combined permutations $p\leftrightarrow q$ and 
$a\leftrightarrow b$. These equations are similar to the unperturbed cluster 
equations in Ref. \cite{lindgren-85}. There is, however, a major difference 
from the all-order equations of unperturbed cluster operators used in other 
works \cite{blundell-89a,blundell-89b} which use antisymmetized cluster 
operators. Here, all order is the same as the linearized coupled-cluster
theory and in the antisymmetized representation $t_{ab}^{pq}=-t_{ba}^{pq}$.
For our work, which is based on diagrammatic evaluation of angular factors, 
the representation without anti-symmetrization is preferable as it follows 
directly from the diagrams. To solve Eq. (\ref{lin_s}) and (\ref{lin_d}), the 
expressions of $\bm{\tau}_{ab}^{pq}$ are separated into multipole
components and the angular factor of each terms in the two equations are
given in the Appendix.


\subsection{One-valence systems}

  From Eq. (\ref{pnc_grnd}), we may write the perturbed eigenvalue equation of 
the one-valence system as
\begin{equation}
  \left ( H^{\rm DC} + \lambda \helec\cdot\mathbf{I} \right) | 
  \widetilde{\Psi}_v \rangle = E_v| \widetilde{\Psi}_v \rangle.
  \label{pnc_1v}
\end{equation}
As defined earlier, $E_v$ is the energy of the one-valence system. Similar to 
the closed-shell case, the perturbed wavefunction in CC theory is 
\begin{equation}
  | \widetilde{\Psi}_v \rangle = e^{T^{(0)}}\left[ 1 
    + \lambda \pt \cdot\mathbf{I} \right] \left[ 1
    + S^{(0)} +\lambda \ps \cdot\mathbf{I} \right] |\Phi_v \rangle,
  \label{psiptrb1v}
\end{equation}
where $\ps$ is the perturbed CC operator of the one-valence part. The 
diagrammatic representation of the valence single and double perturbed
cluster operators $\pso$ and $\pst$, respectively, are shown in 
Fig. \ref{pert_cc_op}. These are topologically equivalent to the diagrams
of the closed-shell operators $\pto$ and $\ptt$, however, with
one of the core lines rotated. The Eq. (\ref{psiptrb1v}) in terms of the 
PRCC wavefunction is 
\begin{eqnarray}
  &&\left( H + \lambda\helec \cdot\mathbf{I} \right)
  e^{T^{(0)}} \left[ 1 + \lambda \pt \cdot\mathbf{I} \right]
  \left[ 1 + S^{(0)} \right . \nonumber \\
 && \left . + \lambda\ps\cdot\mathbf{I} \right] |\Phi_v \rangle 
  = E_v e^{T^{(0)}} \left[ 1 + \lambda \pt\cdot\mathbf{I} \right]
    \left[ 1 + S^{(0)} \right .  \nonumber \\
 && \left. + \lambda \ps\cdot\mathbf{I} \right] |\Phi_v \rangle.
\end{eqnarray}
To derive the $\ps$ equations, like in the closed-shell case, project the 
above equation  on $e^{-T}$ and retain the terms linear in $\lambda$. For 
further simplification, use normal-ordered form of the Hamiltonian, which for 
the one-valence system is $H_{\rm N} = H - \langle\Phi_v|H|\Phi_v\rangle$. 
After these sequence of operations, and retaining only the electronic part
like in the closed-shell case, the eigen value equation is modified to 
\begin{eqnarray}
  \left[ \bar H_{\rm N}\ps  + \bar H_{\rm N}\pt  ( 1 + S ) +
    \bar {\mathbf{H}}_{\rm elec}^{\rm NSD} ( 1 + S ) \right] |\Phi_v \rangle
                          \nonumber \\
  =\left[ \Delta E_v \ps + \Delta E_v \pt ( 1 + S ) \right]|\Phi_v \rangle,
  \label{deltae1v}
\end{eqnarray}
where $\Delta E_v = E_v - \langle\Phi_v|H|\Phi_v\rangle$, is the correlation
energy of the one-valence system. Projecting Eq. (\ref{deltae1v}) with the
excited determinants $\langle\Phi^p_v|$ and $\langle\Phi^{pq}_{va}|$, we get
the perturbed CC equations of the singles and doubles respectively,
in the form
\begin{subequations}
\begin{eqnarray}
  && \langle \Phi^p_v |\{ \contraction{}{H}{_{\rm N}}{S}\bar{H}_{\rm N}
     \mathbf{S}^{(1)} \} + \{ \contraction{}{H}{_{\rm N}}{S}\bar{H}_{\rm N}
     \mathbf{T}^{(1)} \} + \{ \contraction{}{H}{_{\rm N}}{T}
     \contraction[1.5ex]{}{V}{_{\rm N}T^{(1)}}{S}\bar{H}_{\rm N}
     \mathbf{T}^{(1)}S^{(0)}\} + \bar{\mathbf{H}}_{\rm elec}^{\rm NSD} 
                   \nonumber \\
   && + \{ \contraction{}{H}{_{\rm elec}^{\rm NSD}}{S}
     \bar{\mathbf{H}}_{\rm elec}^{\rm NSD}{S}^{(0)} \}|\Phi_v \rangle =
  \Delta E_v \langle \Phi^p_v | \pso|\Phi_v \rangle,
  \label{ccsptrb1v1}
\end{eqnarray}
\begin{eqnarray}
  && \langle \Phi^{pq}_{vb} |\{ \contraction{}{H}{_{\rm N}}{S}\bar{H}_{\rm N}
     \mathbf{S}^{(1)} \} + \{ \contraction{}{H}{_{\rm N}}{S}\bar{H}_{\rm N}
     \mathbf{T}^{(1)} \} + \{ \contraction{}{H}{_{\rm N}}{T}
     \contraction[1.5ex]{}{V}{_{\rm N}T^{(1)}}{S}\bar{H}_{\rm N}
     \mathbf{T}^{(1)}S^{(0)}\} + \bar{\mathbf{H}}_{\rm elec}^{\rm NSD} 
                               \nonumber \\
  &&  + \{ \contraction{}{H}{_{\rm elec}^{\rm NSD}}{S}
     \bar{\mathbf{H}}_{\rm elec}^{\rm NSD}{S}^{(0)} \}|\Phi_v \rangle =
  \Delta E_v \langle \Phi^{pq}_{vb} | \pst|\Phi_v \rangle.
  \label{ccsptrb1v2}
\end{eqnarray}
\label{prcc_eqn}
\end{subequations}
Where, we have used the relations
\begin{equation}
  \langle \Phi^p_v | \pt |\Phi_v \rangle = 0,
  \text{ and }
  \langle \Phi^p_v| \pt S| \Phi_v \rangle = 0,
\end{equation}
as $\pt$, being the cluster operator of closed-shell sector, does not
contribute to the CC equation of $\pso$ and $\pst$. In terms of specific 
components, as in Eq. (\ref{lin_s}), the single excitation perturbed 
cluster amplitude equation is 
\begin{eqnarray}
   (\epsilon_v + &\Delta E_v& - \epsilon_p) \bm{\tau}^p_v = \bm{h}_{pv} 
      + \sum_q \bm{h}_{pq} t^q_v  -\sum_b \bm{h}_{bv} t^p_b 
                  \nonumber \\
  && + \sum_{bq}\bm{h}_{bq}\widetilde t^{qp}_{bv} 
     + \sum_{bq}{\widetilde g}_{bpqv} \bm{\tau}^q_b 
     + \sum_{bqr}{\widetilde g}_{bpqr} \bm{\tau}^{qr}_{bv}  \nonumber \\
  && -\sum_{bcq} g_{bcqv} \bm{\widetilde\tau}^{qp}_{bc}.
  \label{lin_s1}        
\end{eqnarray}
This is Eq. (\ref{lin_s}) with two important modifications. First, the 
valence orbital $v$ replaces the orbital $a$ and second, the single particle
energy of $v$ is modified to include the correlation energy. Following 
similar modifications, the equation of the double excitation cluster 
amplitudes is 
\begin{eqnarray}
  (\epsilon_v + &\Delta E_v& + \epsilon_b - \epsilon_p - \epsilon_q) 
     \bm{\tau}^{pq}_{vb} = \left ( \sum_r\bm{h}_{pr} t^{rq}_{vb}
     -\sum_c\bm{h}_{cv}t^{pq}_{cb} \right .        \nonumber \\
     && +\sum_{r} g_{pqrb} \bm{\tau}^r_v  -\sum_{c} g_{cqvb} \bm{\tau}^p_c 
  +\sum_{rc} g_{pcvr} \bm{\widetilde\tau}^{rq}_{cb} 
                  \nonumber \\
  && \left .  -\sum_{rc} g_{pcrb} \bm{\tau}^{rq}_{vc}  
      -\sum_{rc} g_{cpvr} \bm{\tau}^{rq}_{cb}  \right ) + 
     \left ( \begin{array}{c}
              p\leftrightarrow q \\
              v\leftrightarrow b 
             \end{array} \right ) 
                  \nonumber \\
  &&  + 
  \sum_{rs} g_{pqrs} \bm{\tau}^{rs}_{vb}
      +\sum_{cd} g_{cdvb} \bm{\tau}^{pq}_{cd}.
  \label{lin_s2}                         
\end{eqnarray}
Angular factors obtained from the closed-shell sector, with replacement of 
$a$ by $v$, can be used to rewrite the cluster equations in terms specific
multipole components from these equations. With these modifications, the same
methods used to solve the $T^{(1)}$ are used to calculate the $S^{(1)}$
cluster amplitudes. The fore going derivations shows that PRCC equations of
one-valence sector are not very different from the closed-shell equations.
But the same rationale does not apply to two-valence PRCC equations. For
two-valence systems, even at the level of RCC, require due considerations on
the nature of model space. Properties calculations from the RCC wave 
functions is another level of sophistication, these and related issues on RCC 
calculations of two-valence systems is explored in one of our recent works
\cite{mani-11a}.


\section{$E1_{\rm elec}^{\rm NSD}$ calculation}
\label{e1pnc_elec}

In this section, as prelude to the calculation of $E1_{\rm elec}^{\rm NSD}$ 
from the PRCC states, the details of $ E1_{\rm PNC}^{\rm NSD}$ calculation 
with MBPT is discussed at the beginning. The MBPT calculation, however, is in 
the hyperfine states and is equivalent to the PRCC expressions of 
$E1_{\rm elec}^{\rm NSD}$ after coupling with the nuclear states.


\subsection{From MBPT wavefunction}

\begin{figure}[h]
\begin{center}
  \includegraphics[width = 8.0cm]{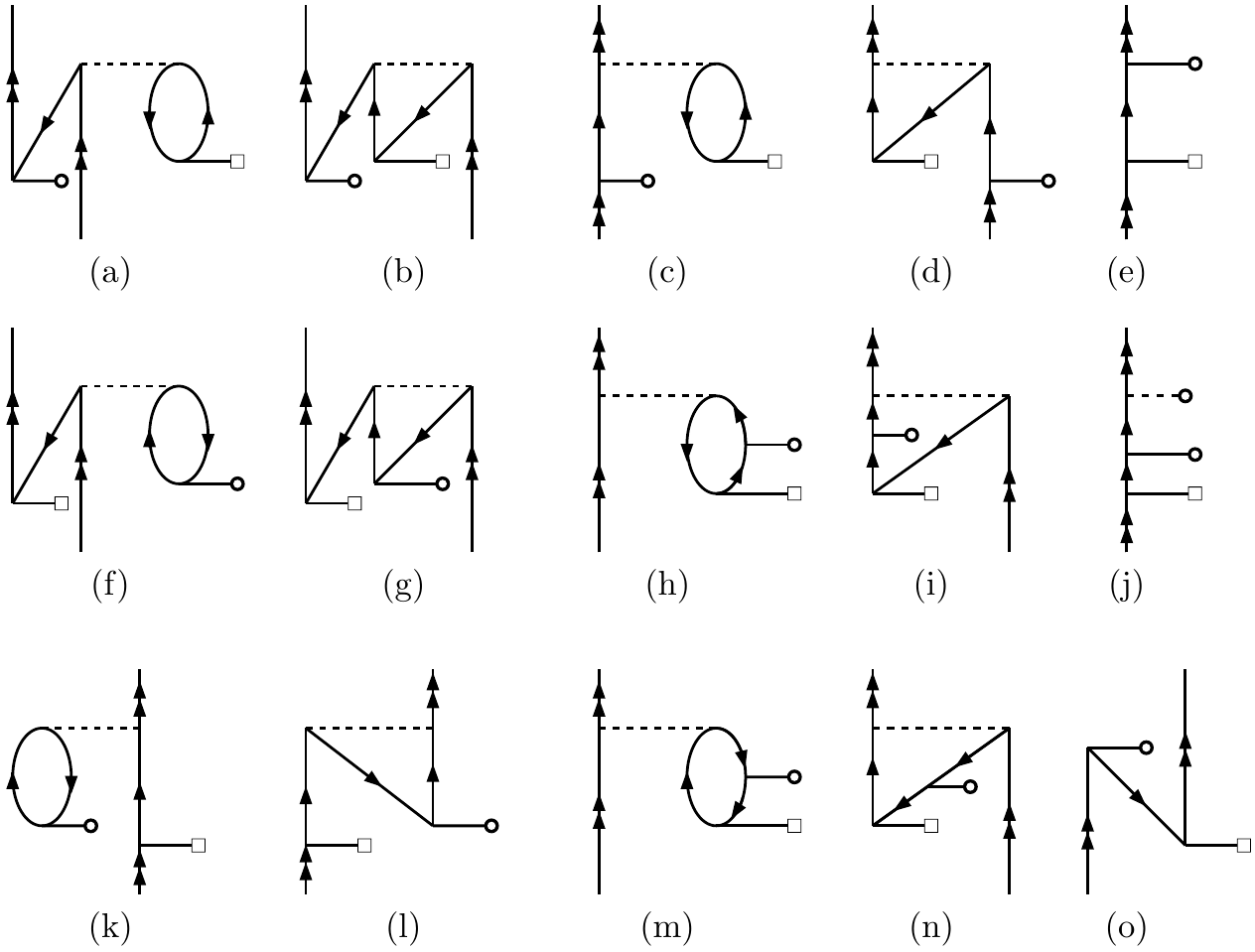}
  \caption{$E1_{\rm elec}^{\rm PNC}$ diagrams which arise from  the MBPT terms
           $\mathbf{D}\Omega^{(0)}_{v,1}$  and 
           ${\Omega^{(1)}_{v,0}}^\dagger \mathbf{D} \Omega^{(0)}_{v,1}$. These
           are the first and third terms of the Eq. (\ref{e1pncmbpt2}). 
           In the diagrams interaction lines terminated with circle and 
           rectangle represent $\mathbf{D} $ and $H_{\rm elec}^{\rm PNC}$,
           respectively.}
  \label{e1pnc_mbpt}
\end{center}
\end{figure}

  To calculate the $H_{\rm PNC}^{\rm NSD} $ induced electric dipole transition 
amplitude $E1_{\rm PNC}^{\rm NSD}$ with MBPT, we use the wavefunction in  
Eq. (\ref{pert_state_mbpt}). For the purpose of using MBPT calculations as
the basics of analyzing the PRCC, it is sufficient to consider the total 
wave operator 
\begin{equation}
  \Omega  = \Omega_{v, 0}^{(1)} + \Omega_{v, 1}^{(0)}.
\end{equation}
The PNC induced $E1$ transition amplitude is then
\begin{eqnarray}
  E1_{\rm PNC}^{\rm NSD} & = & \langle\Phi_w\red\left[1 + \Omega_{w,0}^{(1)} +
              \lambda \Omega^{(0)}_{w, 1}\right]^\dagger\mathbf{D}
                                               \nonumber \\
              &&\left[ 1 + \Omega^{(1)}_{v,0} + \lambda \Omega^{(0)}_{v,1} 
              \right] \red\Phi_v \rangle.
\end{eqnarray}
The expressions arising from these terms contain all the symmetry information
to describe the properties of $T^{(1)}$. Consider terms linear in $\lambda$,
\begin{eqnarray}
   E1_{\rm PNC}^{\rm NSD} & = & \langle \Phi_w \red \mathbf{D} 
      \Omega^{(0)}_{v,1} + {\Omega^{(0)}_{w,1}}^\dagger \mathbf{D}      +
                                     \nonumber \\
      &&{\Omega^{(1)}_{v,0}}^\dagger \mathbf{D} \Omega^{(0)}_{v,1} +
       {\Omega^{(0)}_{w,1}}^\dagger \mathbf{D}\Omega^{(1)}_{v,0}
       \red\Phi_v\rangle. 
       \label{e1pncmbpt2}
\end{eqnarray}
This is the MBPT expression of $ E1_{\rm PNC}^{\rm NSD}$, which has one order 
each of $H_{\rm PNC}^{\rm NSD}$ and residual Coulomb interactions. 
The diagrams which arises from $\mathbf{D} \Omega^{(0)}_{v,1} $ and 
${\Omega^{(1)}_{w,0}}^\dagger \mathbf{D} \Omega^{(0)}_{v,1}$  are shown in 
Fig. \ref{e1pnc_mbpt}. A similar set of diagrams, which arise from 
${\Omega^{(0)}_{w,1}}^\dagger \mathbf{D} $ and 
${\Omega^{(0)}_{w,1}}^\dagger \mathbf{D}\Omega^{(1)}_{v,0} $, are obtained from
interchanging $ H_{\rm PNC}^{\rm NSD}$ and $\mathbf{D}$ vertices in the 
diagrams.  
%
%
Besides providing insights on the nature of perturbed cluster operators, 
detailed MBPT calculations serve another valuable purpose. These serve as
reference calculations to validate the PRCC implementations.


\subsection{From PRCC wavefunction}

\begin{figure}[h]
\begin{center}
  \includegraphics[width = 8.0cm]{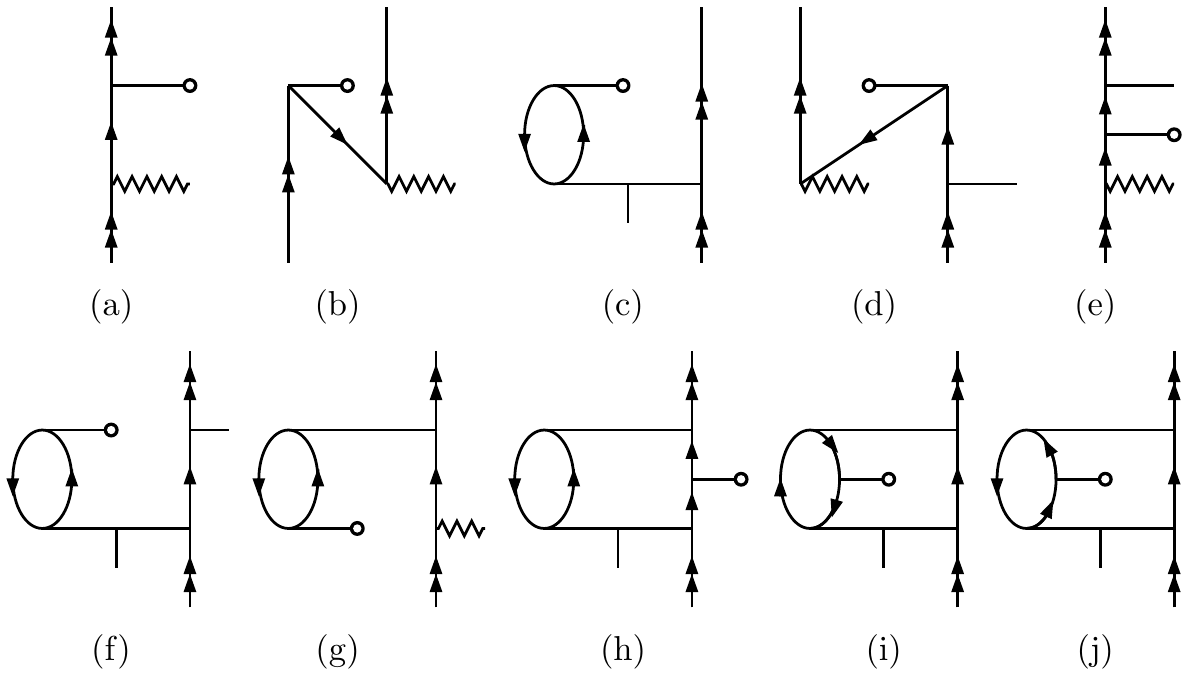}
  \caption{Few of the leading order PRCC diagrams which contribute to the 
           $E1_{\rm elec}^{\rm PNC}$ of one-valence atoms.}
  \label{e1pnc_cc}
\end{center}
\end{figure}

     Like in MBPT, from the expression of the PRCC wave function in 
Eq. (\ref{psiptrb1v}), the $H_{\rm PNC}^{\rm NSD} $ induced $E1$ transition 
amplitude is
\begin{eqnarray}
   E1_{\rm PNC}^{\rm NSD} && =  \langle \Phi_w \red {e^T}^\dagger 
       \left[ 1 + \lambda \pt\cdot\mathbf{I}  \right]^\dagger
       \left[ 1 + S + \lambda \ps\cdot\mathbf{I} \right]^\dagger \mathbf{D}
                          \nonumber \\
       && \times e^T \left[ 1 + \lambda \pt\cdot\mathbf{I} \right]
       \left[ 1 + S + \lambda \ps\cdot\mathbf{I}\right] \red \Phi_v \rangle.
\end{eqnarray}
Retain terms linear in $\lambda$ as the remaining are zero from parity 
selection rule. Considering only the electronic component, like in the 
calculation of the cluster amplitudes $\pt$ and $\ps$, define 
$E1_{\rm elec}^{\rm NSD} $ as the $H_{\rm elec}^{\rm NSD}$ induced $E1$ 
amplitude in the electronic space. From the commutation relation of the 
cluster-operators, the expression is reduced to
\begin{eqnarray}
  E1_{\rm elec}^{\rm NSD}& = & \langle \Phi_w \red \bar{\mathbf{D}} 
     \left[ \pt + \ps + \pt S \right]+ \left[ \pt + \ps  \right. 
                                              \nonumber \\ 
              && \left. + \pt S \right]^\dagger \bar{\mathbf{D}} +
                 S^\dagger\bar{\mathbf{D}} \left[ \pt + \ps + \pt S 
                  \right] \nonumber \\
              && + \left[ \pt + \ps + \pt S \right]^\dagger
                 \bar{\mathbf{D}} S \red \Phi_v \rangle,
\end{eqnarray}
where $\bar{\mathbf{D}} = {e^T}^\dagger \bm{\mathbf{D}} e^T,$ is the dressed 
electric dipole operator. It is evident that $\bar{\mathbf{D}}$ is a 
non-terminating series of the closed-shell cluster operators. It is 
non-trivial to incorporate $T$  to all orders in numerical computations. For 
this reason $\bar{\mathbf{D}}$ approximated as
\begin{equation}
  \bar{\mathbf{D}} \approx \mathbf{D} + \mathbf{D} T^{(0)} + 
      {T^{(0)}}^\dagger \mathbf{D} + {T^{(0)}}^\dagger \mathbf{D} T^{(0)}.
\end{equation}
This captures all the important contributions arising from the 
core-polarization and pair-correlation effects. Terms not included in 
this approximation are terms which are third and higher order in $T^{(0)}$.
The  expression used in our calculations is then
\begin{eqnarray}
   E1_{\rm elec}^{\rm NSD}&  \approx & \langle\Phi_w\red 
               \mathbf{D} \pt + {T^{(0)}}^\dagger \mathbf{D} \pt + 
               {\pt}^\dagger \mathbf{D} T^{(0)}  
                                                 \nonumber \\
            && +{\pt}^\dagger \mathbf{D} + \mathbf{D} \pt S^{(0)} 
               + {\pt}^\dagger {S^{(0)}}^\dagger \mathbf{D} 
                                                 \nonumber \\
            && +{S^{(0)}}^\dagger \mathbf{D} \pt
               + {\pt}^\dagger \mathbf{D} S^{(0)} + \mathbf{D} \ps
               +{\ps}^\dagger \mathbf{D}                      \nonumber \\
            && + {S^{(0)}}^\dagger \mathbf{D} 
               \ps + {\ps}^\dagger \mathbf{D} S^{(0)} \red\Phi_v\rangle.
  \label{e1pnccc1v}
\end{eqnarray}
The Eq. (\ref{e1pnccc1v}), as mentioned earlier, includes term up to second 
order in cluster amplitudes. From our previous study of properties calculations 
\cite{mani-10}, we conclude that the contributions from the higher order 
are negligible. Selected diagrams from the leading-order and next to leading
order terms of Eq. (\ref{e1pnccc1v}) are shown in the Fig. \ref{e1pnc_cc}.
%
%
%
%


\section{Coupling with nuclear spin}
\label{couple_nsd}

  The PRCC diagrams of $E1_{\rm elec}^{\rm NSD}$ in Fig. \ref{e1pnc_cc} are, 
as explained earlier, defined in the electronic subspace. To calculate
$E1_{\rm PNC }^{\rm NSD}$, the nuclear spin part of the operator is coupled 
with $E1_{\rm elec}^{\rm NSD}$. The $E1_{\rm PNC}^{\rm NSD}$ diagrams must 
then include a vertex which operates on the nuclear spin states. This 
is evident from the example MBPT diagram in Fig. \ref{pcc_mbpt}$(a)$. 
Although complicated, the coupling with the nuclear part of the operator 
require only angular integration, where as  the evaluation in the electronic 
space involves radial integrals. In our calculations, we use diagrammatic
to carry out the angular integrals and as mentioned earlier, we follow the
conventions of Lindgren and Morrison \cite{lindgren-85}.

  To elaborate on the coupling schemes, we discuss specific cases involving 
single and double PRCC operators. For the former, we select the topologically
simplest, and for the later, one of the more complicated one. So that, these
represent the range of topology and complexity of the diagrams. In each of 
these the main objective is to reduce the electronic part to a form common to 
a set of diagrams. And then, combine with the nuclear part.
\begin{figure}[h]
\begin{center}
  \includegraphics[width = 8.2cm]{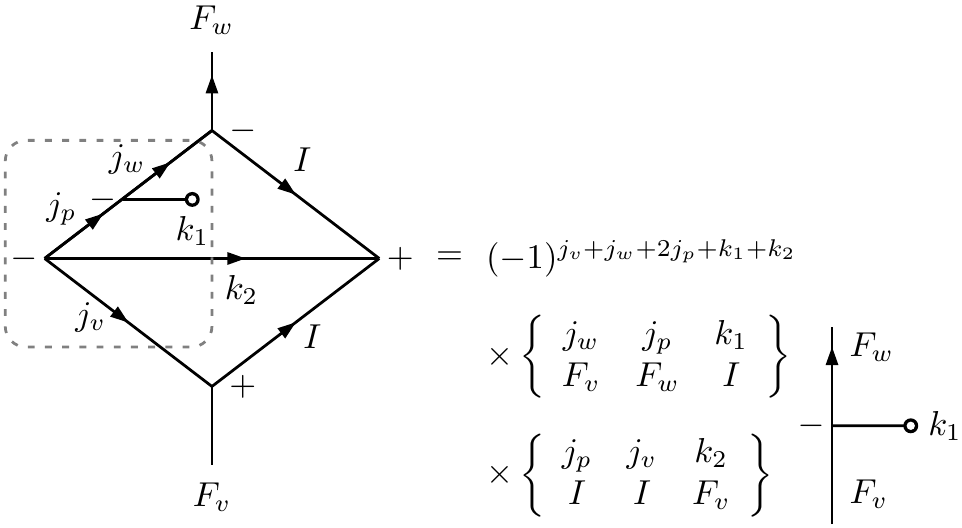}
  \caption{Angular factor calculation of the $E1_{\rm PNC}^{\rm NSD}$ 
           diagram shown in Fig. \ref{e1pnc_cc}(a) and it arises from the 
           term $D S^{(1)}_1$. The portion within the rectangle in dashed line
           indicate the electronic part. Remaining portion is from the nuclear 
           spin and hyperfine coupling.}
  \label{e1pnc_d5}
\end{center}
\end{figure}


\subsection{Singles PRCC operator }

 To define the coupling of singles diagrams, either $ \pto$ or $\pso$,
with nuclear spin part consider the diagram in Fig. \ref{e1pnc_cc}(a).  It 
represents the Dirac-Fock contribution and is perhaps the simplest diagram. 
But it is the most dominant and naturally the most important. It is one of the
diagrams arising from $ D\pso$ in the PRCC calculations. The angular 
diagram for the electronic part of this diagram has same topology as in 
Fig. \ref{e1pnc_cc}(a). However, the orbitals lines are replaced by angular 
momentum lines and arrows on appropriate lines to represent phase factors.
The angular momentum diagram, which includes coupling with the nuclear spin, is 
shown in Fig. \ref{e1pnc_d5}. The electronic part is the portion enclosed 
with the dashed line rectangle.

   In the diagram, $j_v$  and $j_w$ are the total angular momentum of the 
single particle states of the initial and final one-valance atomic states, 
respectively. The line $k_1$ represents a rank one multipole line and denotes 
the angular part of the dipole operator. The multipole line $k_2$ is the all 
important representation of $H_{\rm PNC}^{\rm NSD}$, it is rank one multipole 
in electronic part as well as the nuclear part. In the electronic part, it 
represents the angular part of operator $\bm\alpha $ and produces a transition 
from $j_v$ to $j_p$. Here, $j_p$ is the total angular  momentum of the 
intermediate single particle state. On the other hand, in the nuclear sector, 
$k_2$ represents the operator $\mathbf{I}$. Since it is diagonal operator of 
the nuclear spin state $|I\rangle $, the $k_2$ line does not change the 
nuclear spin. 

  The last step in the angular momentum diagram representation is to coupled
the electron and nuclear momenta to form the total angular momentum of the 
hyperfine states. The initial and final hyperfine states are $|F_v\rangle$ and
$|F_w\rangle$, respectively. The lines marked as $F_v$ and $F_w$ represent the
angular momenta of the hyperfine states in the diagram. Using Wigner-Eckert
theorem, we can write algebraic equivalent of the diagram as
\begin{eqnarray}
   \langle F_wm_w| D\pso\cdot\mathbf{I}|F_vm_v\rangle & = & (-1)^{F_w - m_w}
       \left ( \begin{array}{ccc}
                 F_w & 1 & F_v \\
                -m_w & q & m_v 
               \end{array} \right ) \nonumber \\
       && \times \langle F_w\red D_{\rm eff}\red F_v\rangle,
  \label{deff_sing}
\end{eqnarray}
where $D_{\rm eff} = D\pso\cdot\mathbf{I}$ is a rank one operator, $m_i$ are 
the hyperfine magnetic quantum numbers and $q$ is the component of 
$D_{\rm eff}$. The phase factor and $3j$-symbol in the above expression are 
the free lines in the right hand side of the diagrammatic equation in 
Fig. \ref{e1pnc_d5}. Remaining expression on the right hand side, phase factor 
and $6j$-symbols, is the $m$ independent angular component of 
$\langle F_w\red D_{\rm eff}\red F_v\rangle $.

\begin{figure}[h]
\begin{center}
  \includegraphics[width = 7.0cm]{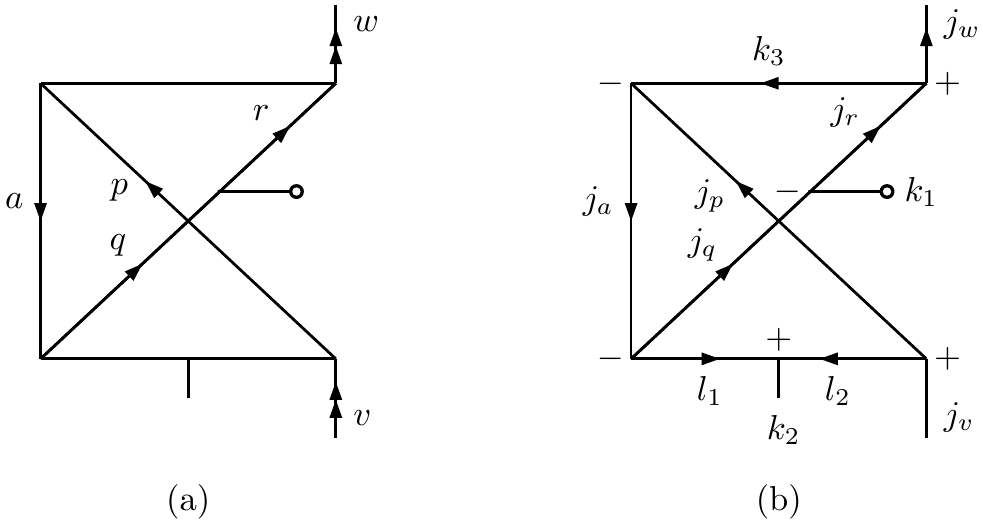}
  \caption{Angular momentum diagram representation of a PRCC 
           $E1_{\rm elec}^{\rm NSD}$ diagram. (a) Exchange of the diagram 
           shown in Fig. \ref{e1pnc_cc}(j). (b) The corresponding angular
           momentum diagram.}
  \label{e1pnc_d1}
\end{center}
\end{figure}


\subsection{Doubles PRCC operator }

For the coupling with nuclear spin involving either $\pto$ or 
$\pso$, consider a more complicated diagram as  shown in 
Fig. \ref{e1pnc_d1}(a). It is the exchange of diagram in 
Fig. \ref{e1pnc_cc}(j) and arises from the term ${T_2^{(0)}}^{\dagger}D\pst$ 
in the PRCC expression of $E_{\rm elec}^{\rm NSD}$. The angular momentum 
diagram is shown in Fig. \ref{e1pnc_d1}(b) and note that the angular momentum 
representation of $\pst$  is as described in Section. \ref{pert_cc_fn}. The
labels of the lines are different from the one in Fig. \ref{pcc_mbpt1}, 
however, these are dummy labels and same selection rules apply to $l_1$ and 
$l_2$, and $k_2$ represents a rank one multipole line. It has the same role 
in the electronic sector as $k_2$ of the diagram in Fig. \ref{e1pnc_d5}. 
\begin{figure}[h]
\begin{center}
  \includegraphics[width = 3.8cm]{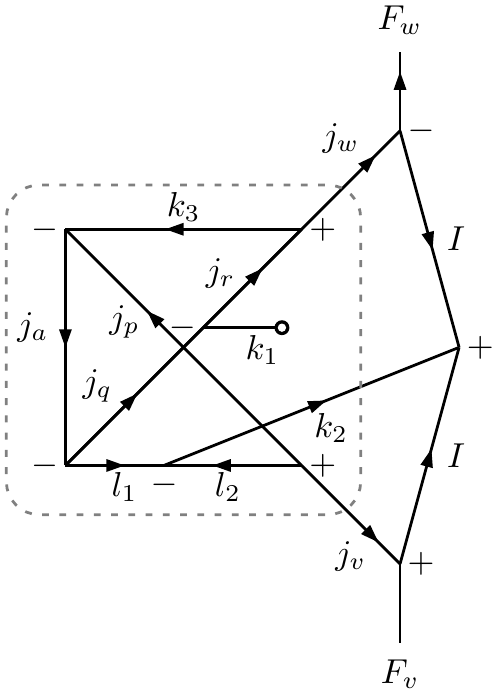}
  \caption{Angular momentum diagram of $E1_{\rm PNC}^{\rm NSD}$ in the 
           hyperfine states involving double excitation cluster operator
           $S_2^{(1)}$. The diagram arises from the term 
           ${S_2^{(0)}}^{\dagger}\mathbf DS_2^{(1)}$ and the portion within
           the rectangle in dashed lines indicate the portion arising from 
           the electronic sector.}
  \label{e1pnc_d2}
\end{center}
\end{figure}

 To demonstrate the non-trivial angular integration in the calculations with 
hyperfine atomic states, the angular momentum diagram of 
$\langle F_wm_w|{T_2^{(0)}}^{\dagger}D\pst\cdot\mathbf{I}|f_vm_v\rangle$, of 
the diagram in Fig. \ref{e1pnc_d1}, is shown in Fig. \ref{e1pnc_d2}. The 
portion of the diagram within the rectangle in dashed-line is the angular 
momentum part of the electronic sector and except for the topological rotation
of the ($l_1$, $l_2$, $k_2$ ) vertex,  it is identical to the diagram
in Fig. \ref{e1pnc_d1}(b).
\begin{figure}[h]
\begin{center}
  \includegraphics[width = 7.8cm]{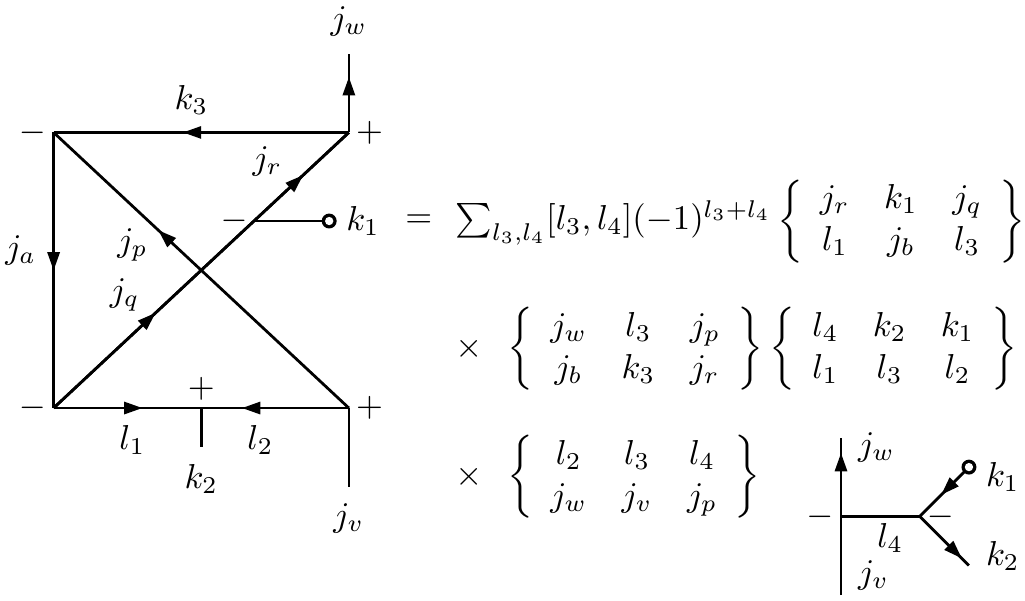}
  \caption{Angular factor reduction in the electronic sector. The angular 
           momentum diagram on the left hand side is the portion within 
           rectangle in Fig. \ref{e1pnc_d2}. On the right hand side, the
           angular momentum diagram with free lines represents the effective
           part coupled to the nuclear spin.}
  \label{e1pnc_d3}
\end{center}
\end{figure}
The evaluation of the angular integral of the electronic sector, for the 
example considered, is shown in Fig. \ref{e1pnc_d3}. Important point to be
observed is the structure of the free part in the right hand side
of the diagrammatic equation in Fig. \ref{e1pnc_d3}. Although the multipole
lines of $D$ and $\bm{\alpha}$, $k_1 $ and $k_2 $ respectively, are coupled 
to an effective multipole $l_4$, the $k_1 $ and $k_2 $ are present as free 
lines. The effective multipole $l_4$ operates on $j_v$ and transforms it to
$j_w$.  In terms of Wigner-Eckert theorem, the diagram is equivalent to 
\begin{eqnarray}
   && \langle j_wm_w| \sum_{l_4}\left \{ {T_2^{(0)}}^{\dagger}\bfd\pso 
        \right\}^{l_4} |j_vm_v\rangle  =  (-1)^{j_w - m_w}  \nonumber \\
  && \times \left ( \begin{array}{ccc}
                       j_w & l_4 & j_v \\
                      -m_w & q & m_v 
                    \end{array} \right ) \langle j_w\red
      \left \{{T_2^{(0)}}^{\dagger} \bfd\pso \right\}^{l_4} \red j_v\rangle,
  \label{deff_dbl_elec}
\end{eqnarray}
where $\{\ldots\}^{l_4}$ represents coupling of rank one tensor operators 
$\mathbf{D} $ and $\ps$ to an operator of rank $l_4$. This coupling is a 
structure common to any PRCC term of $E1_{\rm elec}^{\rm NSD}$. That is, for 
any term, the angular integral in the electronic sector is reducible to a form
where the free lines is similar to the one on the right hand side of
Fig. \ref{e1pnc_d3}. 
\begin{figure}[h]
\begin{center}
  \includegraphics[width = 8.2cm]{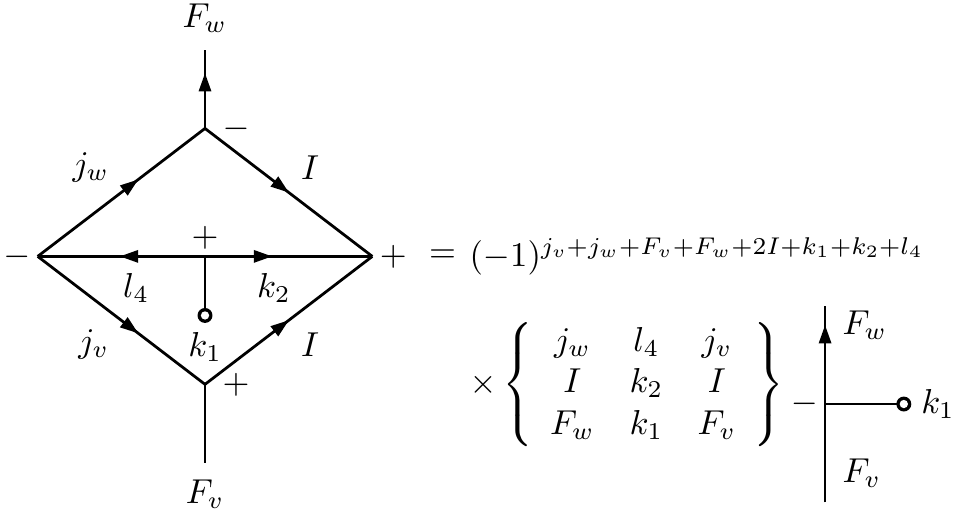}
  \caption{ Angular factor reduction of the coupling between the electronic 
            sector with the nuclear spin of a diagram arising from
           ${S^{(0)}}^\dagger_2 D S^{(1)}_2$. The electronic sector of the 
           diagram is shown in Fig. \ref{e1pnc_d3}.}
  \label{e1pnc_d4}
\end{center}
\end{figure}
From the triangular condition, $l_4 = 0, 1, 2$ are the allowed values, 
however, what values of $l_4$ contribute depends on $j_v$ and $j_w$. For 
example, $l_4=0,1$ contribute in the PNC $6\;^2S_{1/2}\rightarrow 7\;^2S_{1/2}$ 
transition of atomic Cs \cite{wood-97}, where as only $l_4=2$  contributes
to the proposed PNC $6\;^2S_{1/2} \rightarrow 5\; ^2D_{5/2}$  transition in 
Ba$^+$ \cite{fortson-93}.

  The form of the free lines in the Fig. \ref{e1pnc_d3} require one due
consideration while combining with the nuclear part. It is the multipole line
$k_2$, inherited from $\pt$ or $\ps $, which couples with the operator 
$\mathbf{I}$ to form a scalar operator. The diagrammatic representation is 
shown on the left side of Fig. \ref{e1pnc_d4}.  After evaluation, it reduces to
a $9j$-symbol and free line part. From Wigner-Eckert theorem, the matrix
element in the hyperfine states is
\begin{eqnarray}
   && \sum_{l_4}\langle F_wm_w| \left \{ \left [ {T_2^{(0)}}^{\dagger}\bfd\pso 
        \right]^{l_4} \mathbf{I}\right \}^1 |F_vm_v\rangle  =  (-1)^{F_w - m_w}  \nonumber \\
  && \times \left ( \begin{array}{ccc}
                       F_w & 1 & F_v \\
                      -m_w & q & m_v 
                    \end{array} \right ) \langle F_w\red
      D_{\rm eff}\red F_v\rangle,
  \label{deff_dbl}
\end{eqnarray}
where
\begin{equation}
  D_{\rm eff} = \left \{ \left [ {T_2^{(0)}}^{\dagger}\bfd\pso 
        \right]^{l_4} \mathbf{I}\right \}^1 , 
\end{equation}
is the effective dipole operator in the hyperfine states. It is of the same 
form as the effective operator in Eq. (\ref{deff_sing}). However, the 
coupling sequence in Eq. (\ref{deff_dbl}) is general and applies to all
the terms in the $E1_{\rm PNC}^{\rm NSD}$.


\section{Validation of PRCC}
\label{validation_prcc}

  The RCC method described so far involves intricate but tractable angular 
momentum coupling and most of the calculations are in the configuration space 
of reduced dimension, namely of the electrons. Considering the complexity of 
the method, it is desirable to validate the method with few selected terms 
before a full scale implementation. Here, we present a method of validation by
comparing with dominant third order MBPT diagrams. This is possible as we
solve the PRCC equations iteratively with the first order MBPT wave functions
as the initial guess. In particular, we evaluate the $E1^{\rm NSD}_{\rm PNC}$
of the transition $6s \rightarrow 7s $ in $^{133}$Cs. 

\begin{figure}[h]
\begin{center}
  \includegraphics[width = 6.0cm]{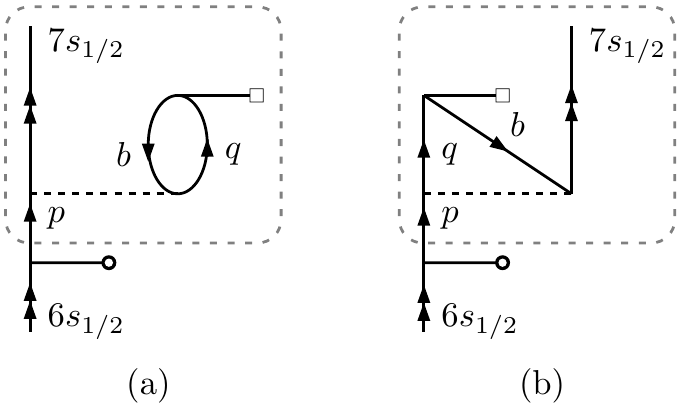}
  \caption{MBPT $E1_{\rm PNC}^{\rm NSD}$ diagrams used in the validation of the
           the singles PRCC amplitudes $S_1^{(1)}$.}
  \label{validation_s}
\end{center}
\end{figure}

\subsection{Single excitation operator $S_1^{(1)}$ }

 To check the PRCC equation and angular momentum coupling in the calculations
of $S_1^{(1)}$, consider the term ${S_1^{(1)}}^{\dagger}D$ in 
$E^{\rm NSD}_{\rm PNC}$. Two of the third order MBPT diagrams, which are 
equivalent to the ${S_1^{(1)}}^{\dagger}D$ at the first iteration of the PRCC
equation are shown in Fig. \ref{validation_s}. In the figure, the portion of 
the diagrams within the rectangle of dashed-line arises from the second order 
wave operator ${\Omega_{7s,1}^{(1)}}^{\dagger}$ define in Eq. (\ref{omega_11}).
As for the solutions of linearized PRCC equations, the initial guess is
the first order MBPT wave function $\Omega_{7s, 1}^{(0)}$. Single excitation 
cluster amplitudes $\bm{\tau} $ are then calculated iteratively from 
Eq. (\ref{lin_s}). At the first 
iteration of the linearized PRCC equation, the diagram within the dashed 
rectangle in Fig. \ref{validation_s}(a) is equal to the one valence version 
of the diagram in Fig. \ref{psingles}(c). And, the diagram 
Fig. \ref{validation_s}(b) is the exchange counterpart. These MBPT diagrams 
are the dominant ones after the DF and other theories like the PRCC must be
able to reproduce matching results.
\begin{table}[t]
\begin{center}
\caption{Validation of singles PRCC amplitudes. The values listed are in 
         units of $iea_0 \mu{'}_w$. Numbers in the square bracket represent 
         the power of 10.}
\label{table_s1}
\begin{ruledtabular}
\begin{tabular}{ccc}
Orbital ($p$,$q$) & MBPT & PRCC                                     \\
\hline
  $6p_{1/2}$  & $3.80237[-15]$ & $3.80237[-15]$    \\
  $7p_{1/2}$  & $-4.42335[-16]$ & $-4.42335[-16]$  \\
  $8p_{1/2}$  & $1.08456[-16]$ & $1.08456[-16]$    \\
  $9p_{1/2}$  & $1.88997[-16]$ & $1.88997[-16]$    \\
  $10p_{1/2}$ & $8.66134[-17]$ & $8.66134[-17]$   \\
  $11p_{1/2}$ & $2.41491[-18]$ & $2.41491[-18]$   \\
\end{tabular}
\end{ruledtabular}
\end{center}
\end{table}

 For comparison, $E1_{\rm PNC}^{\rm NSD}$ contribution from the two diagrams
in Fig. \ref{psingles} from MBPT and results from the equivalent PRCC 
calculations are listed in Table. \ref{table_s1}. The specific orbital wise 
contributions, for better comparison, of the dominant contributions are listed 
in the table.  It is evident that there is excellent agreement between
the MBPT and PRCC results. 
\begin{figure}[h]
\begin{center}
  \includegraphics[width = 6.0cm]{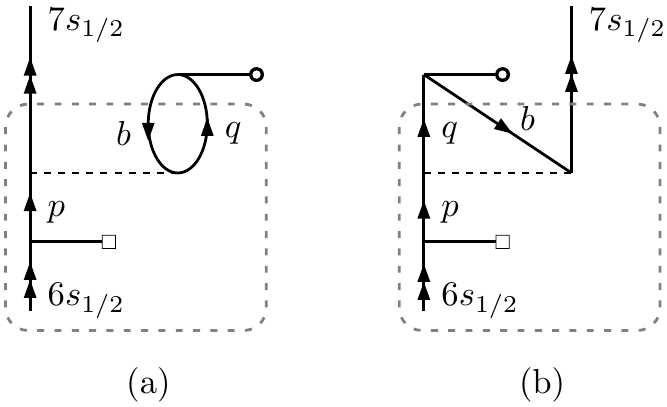}
  \caption{MBPT $E1_{\rm PNC}^{\rm NSD}$ diagrams used in the validation of
           the doubles PRCC amplitudes $S_2^{(1)}$.}
  \label{validation_d}
\end{center}
\end{figure}

\subsection{Double excitation operator $S_2^{(1)}$ }

Like in the case of $S_1^{(1)}$, consider two MBPT diagrams which are
equivalent to the $S_2^{(1)}$ diagram after the first iteration and are
shown in Fig. \ref{validation_d}. In the figure, the portion of the diagrams
within the rectangle in dashed-lines are equivalent to the $S_2^{(1)}$
diagram in Fig. \ref{pdoubles}(a) but adapted for one-valence systems. 
However, there is a major difference from the $S_1^{(1)}$. As 
$H_{\rm PNC}^{\rm NSD}$ is a single particle operator, the double excitation 
cluster operator at one order of $H_{\rm PNC}^{\rm NSD}$ is zero and the 
initial guess value is set to zero. To provide a wider test sample, for the 
double excitation the MBPT diagrams considered are equivalent to the term 
$DS_2^{(1)}$ in PRCC. Another variation is, the two diagrams considered in 
$S_1^{(1)}$ arise from topologically different diagrams, in the present case,
the two diagrams in Fig. \ref{validation_d} arise from the same $S_2^{(1)}$ but
different cluster amplitudes. 
\begin{table}[h]
\begin{center}
\caption{Validation of doubles PRCC amplitudes. The values listed are in 
         units of $iea_0 \mu{'}_w$. Numbers in the square bracket represent 
         the power of 10.}
\label{table_s2}
\begin{ruledtabular}
\begin{tabular}{ccc}
Orbital ($p,q$) & MBPT & PRCC                                   \\
\hline
$6p_{1/2}$ & $6.53322[-17]$ & $6.53322[-17]$  \\
$7p_{1/2}$ & $2.35102[-17]$ & $2.35102[-17]$  \\
$8p_{1/2}$ & $3.20651[-16]$ & $3.20651[-16]$  \\
$9p_{1/2}$ & $2.02539[-15]$ & $2.02539[-15]$  \\
$10p_{1/2}$ & $8.67435[-16]$ & $8.67435[-16]$ \\
$11p_{1/2}$ & $3.02443[-17]$ & $3.02443[-17]$ \\
\end{tabular}
\end{ruledtabular}
\end{center}
\end{table}

The results from the MBPT and PRCC calculations are listed in 
Table. \ref{table_s2}. Here too, the results from the two calculations are
in excellent agreement. Perhaps, it must be mentioned that, the angular
momentum factor calculations of the MBPT and PRCC diagrams are done in 
very different steps. Electronic part of the portion within the rectangle
in dashed-line in Fig. \ref{validation_d} are evaluated to reduce to the
representation of $S_2^{(1)}$ as discussed in Section. \ref{pert_cl_rep}.
MBPT diagrams, on the other hand, may be evaluated without the need to 
associate with an effective operator of specific form.


\section{Conclusions}

  The relativistic coupled-cluster method we have developed to incorporate
a nuclear spin-dependent perturbation is an apt one to calculate nuclear 
spin-dependent parity non-conservation in atoms and ions. The representation
of the cluster amplitude within the electronic sector as a tensor operator
of rank one and coupling with the nuclear spin part at a later stage of 
property calculation leads to simplification of the calculations. Otherwise,
the entire calculation must be done with hyperfine states, which involves
complicated angular momentum couplings at all stages of the calculations. 
The proposed scheme, on the other hand, introduces the nuclear spin coupling
is with a simplified effective operator in the electronic sector. The validity 
of the representation is explicitly tested and verified with selected diagrams.
Although limited in number, the diagrams and terms selected in the sample 
test calculations are varied enough to account for complex as well as subtle 
issues related to the method. Based on the results presented, we conclude that
the method works and in future publications we shall report results of 
sophisticated and large scale calculations using PRCC.


\begin{acknowledgments}
We wish to thank S. Chattopadhyay, S. Gautam, K. V. P. Latha, B. Sahoo and
S. A. Silotri  for useful discussions. The results presented in the paper
are based on computations using the HPC cluster at Physical Research
Laboratory, Ahmedabad.
\end{acknowledgments}


\appendix*
\section{Angular factors of PRCC equation}

 Here, we give the linearized PRCC 
equations of the closed-shell sector, Eqs. (\ref{lin_s}) and (\ref{lin_d}), 
after angular integration. The 
one-valence cluster amplitude equations can be obtained after suitable 
modifications.

\subsection{Single excitation cluster operator}

The angular reduction of each diagram is such that the free parts are reduced 
to the form in Fig. \ref{pert_cc_op}(a). The free part is common to all the 
diagrams and is avoided in the computational implementation. So, solutions of
the PRCC equations are the cluster amplitudes in reduced matrix form. 
For $T_1^{(1)}$ and $S_1^{(1)} $, the rank of the operator, and hence the 
free part, is one. For convenience, the representation of $T_2^{(1)}$ is
redefined as 
\begin{equation}
   \ptt = \sum_{abpq}\sum_{l_1, l_2} \tau_{ab}^{pq}(l_1, l_2) 
       \{\mathbf{C}_{l_1}(\hat{r}_1)\mathbf{C}_{l_2}(\hat{r}_2)  \}^1.
\end{equation}
Like the PRCC operators, we use $t^p_a$ and $t^{pq}_{ab}$ to represent the 
unperturbed single and double excitation amplitudes, respectively. The 
tensor operator structure of $T^{(0)}_2$ is
\begin{equation}
  T^{(0)}_2 = \sum_{abpq}\sum_{k} t_{ab}^{pq}(k)
  \mathbf{C}_k(\hat{r}_1)\cdot\mathbf{C}_{k}(\hat{r}_2).
\end{equation}
For the Slater integrals, the reduced matrix element is 
\begin{equation}
  X_k(abcd) = (-1)^k \langle \kappa_a \red \mathbf{C}^k \red \kappa_c \rangle 
           \langle \kappa_b \red \mathbf{C}^k \red \kappa_d \rangle R_k(abcd), 
\end{equation}
where, $R_k(abcd)$ is the radial part of the Slater integral. With these
definitions, the Eq. (\ref{lin_s}) is written in terms of reduced matrix 
elements and appropriate angular factors.
\begin{widetext}
\begin{eqnarray}
  (\epsilon_a - &\epsilon_p&) \bm{\tau}^p_a = \bm{h}_{pa} 
    + \sum_q \frac{\delta{(j_a, j_q})}{\sqrt{(2 j_a + 1)}} \bm{h}_{pq} t^q_a 
    - \sum_b \frac{\delta{(j_b, j_p})}{\sqrt{(2 j_b + 1)}} \bm{h}_{ba} t^p_b  
    + \sum_{bqk_2}\bm{h}_{bq}\left [ \frac{\delta{(k_1, k_2})}
      {\sqrt{(2 k_1 + 1)}} (-1)^{j_q - j_b + k_1}  t^{qp}_{ba}(k_2) \right .
                 \nonumber \\
  &&-  (-1)^{j_b + j_q + k_1} \left \{ \begin{array}{ccc}
                                                   j_b  & j_q & k_1 \\
                                                   j_a & j_p  & k_2
                                                 \end{array}\right\}  
       t^{qp}_{ab}(k_2) \Bigg ]
    + \sum_{bqk_2} \bm{\tau}^q_b\Bigg [ \frac{\delta{(k_1, k_2})}
      {\sqrt{(2 k_1 + 1)}} (-1)^{j_q - j_b + k_1}   X_{k_2}(bpqa) 
                   \nonumber \\
  &&-  (-1)^{j_b + j_q + k_1} \left \{ \begin{array}{ccc}
                                                    j_a  & j_b & k_2 \\
                                                    j_q & j_p  & k_1
                                                 \end{array}\right\}  
       X_{k_2}(bpaq) \Bigg ] 
    + \sum_{bqrk_2} \sum_{l_1 l_2}\bm{\tau}^{qr}_{ba}(l_1,l_2)
      \Bigg [ \frac{\delta{(k_2, l_2})}{\sqrt{(2 k_2 + 1)}}
      (-1)^{j_q - j_b + j_a + j_p + l_1} 
                             \nonumber \\
  &&\times\;\left \{ \begin{array}{ccc}
                        j_r  & j_a & l_1 \\
                        k_1 & k_2  & j_p
                     \end{array}\right\}   X_{k_2}(bpqr)
    - (-1)^{j_a + j_p + j_b + j_q + l_1}
      \left \{ \begin{array}{ccc}
                  j_b  & k_2 & j_r \\
                  j_p & l_2  & j_q
               \end{array}\right\} 
      \left \{ \begin{array}{ccc}
                        j_r  & j_a & l_1 \\
                        k_1 & l_2  & j_p
                     \end{array}\right\} X_{k_2}(bprq)  \Bigg ]
                     \nonumber \\ 
   && - \sum_{bcqk_2} \sum_{l_1 l_2} X_{k_2}(bcqa)\Bigg [ 
        \frac{\delta{(k_2, l_2})} {\sqrt{(2 k_2 + 1)}} 
      (-1)^{j_a - j_b + j_p + j_q + k_1 + k_2} 
      \left \{ \begin{array}{ccc}
                 j_p  & l_1 & j_c \\
                 k_2 & j_a  & k_1
               \end{array}\right\}  \bm{\tau}^{qp}_{bc}(l_1,l_2)
                     \nonumber \\ 
   && - (-1)^{j_q + j_c + j_a + j_p + k_1 + l_2}
      \left \{ \begin{array}{ccc}
                  k_2  & j_a & j_c \\
                  l_2 & j_q  & j_b
               \end{array}\right\}    
      \left \{ \begin{array}{ccc}
                  j_b  & j_p & l_1 \\
                  k_1 & l_2  & j_a
               \end{array}\right\}  \bm{\tau}^{qp}_{cb}(l_1,l_2) \Bigg ].
\end{eqnarray}
\end{widetext}


\subsection{Double excitation cluster operator}
For the double excitation PRCC operator $T_2^{(1)} $, the common free part of 
the angular factors in Eq. (\ref{lin_d}) is as shown in 
Fig. \ref{pert_cc_op}(b). In the angular factors, the multipoles $k_1$, $k_2$, 
$l_1$ and $l_2$ have the same interpretations as in the singles. Multipoles
$l_3$ and $l_4$ are, however, arise from coupling $(j_a, j_p)$ and 
$(j_b, j_q)$, respectively. The double excitation cluster equation obtained 
from the projection on $\langle\Phi_{ab}^{pq}|$  along with appropriate 
angular factors is 
\begin{widetext}
\begin{eqnarray}
  (\epsilon_a + &\epsilon_b &- \epsilon_p - \epsilon_q) \bm{\tau}^{pq}_{ab}
    (l_1, l_2) = \Bigg [ \sum_{r}(2 l_1 + 1) 
    (-1)^{j_a + j_p + k_1 + l_2} 
    \left \{ \begin{array}{ccc}
               k_1  & j_p & j_r \\
               j_a & l_2  & l_1
             \end{array}\right\}\bm{h}_{pr} t^{rq}_{ab}(l_2)
   - \sum_{c}(2 l_1 + 1)   
                                     \nonumber \\
  && \times\;(-1)^{j_a + j_p + l_1}\left \{ \begin{array}{ccc}
                        k_1  & j_a & j_c \\
                        j_p & l_2  & l_1
                      \end{array}\right\} \bm{h}_{ca} t^{pq}_{cb}(l_2)
  +\sum_{r} (2 l_1 + 1) (-1)^{j_a + j_p + l_1}
    \left \{ \begin{array}{ccc}
               l_2  & j_p & j_r \\
               j_a & k_1  & l_1
             \end{array}\right\} X_{l_2}(pqrb) \bm{\tau}^r_a
                                     \nonumber \\
  && -\sum_{c} (2 l_1 + 1)(-1)^{j_a + j_p + k_1 + l_2}
     \left \{ \begin{array}{ccc}
                l_2  & j_a & j_c \\
                j_p & k_1  & l_1
              \end{array}\right\} X_{l_2}(cqab) \bm{\tau}^p_c
  + \sum_{rc} \frac{1}{(2 l_1 + 1)}
    (-1)^{j_r - j_c + l_1}
                                      \nonumber \\
  &&\times X_{k_2}(pcar)\bm{\tau}^{rq}_{cb}(l_1,l_2) 
    -\sum_{rc}\sum_{l_3 l_4}(2 l_2 + 1) 
     (-1)^{j_c - j_q + k_1 + l_3}
    \left \{ \begin{array}{ccc}
               j_r & l_3  & j_b \\
               j_c & l_4  & j_q \\
               l_1 & k_1  & l_2 
             \end{array}\right\}X_{l_1}(pcar) \bm{\tau}^{rq}_{bc}(l_3,l_4)
                                      \nonumber \\
  && - \sum_{rck_2} \sum_{l_3l_4}(2 l_1 + 1)(2 l_2 + 1) 
     (-1)^{j_a + j_p + j_b + j_q + k_1}   
     \left \{ \begin{array}{ccc}
        k_2  & j_p & j_r \\
        j_a & l_3  & l_1
  \end{array}\right\}             
  \left \{ \begin{array}{ccc}
        l_2  & l_1 & k_1 \\
        l_3 & l_4  & k_2
  \end{array}\right\}
  \left \{ \begin{array}{ccc}
        k_2  & j_c & j_b \\
        j_q & l_2  & l_4
  \end{array}\right\} X_{k_2}(pcrb)  
                                      \nonumber \\
  &&\times \bm{\tau}^{rq}_{ac}(l_3,l_4)
     -\sum_{rck_2} (-1)^{j_c + j_r + l_3}
     \left \{ \begin{array}{ccc}
        j_a  & j_c & k_2 \\
        j_r & j_p  & l_1
  \end{array}\right\} X_{k_2}(cpar)\bm{\tau}^{rq}_{cb}(l_1,l_2), \Bigg ]
  +   \Bigg [ \begin{array}{c}
              p\leftrightarrow q \\
              a\leftrightarrow b 
             \end{array} \Bigg ] 
                                      \nonumber \\
  &&+\sum_{rsk_2}\sum_{l_3 l_4}(2 l_1 + 1)(2 l_2 + 1) 
     (-1)^{j_a + j_p + j_b + j_q + k_1 + k_2 + l_4 + l_2}
      \left \{ \begin{array}{ccc}
        k_2  & j_p & j_r \\
        j_a & l_3  & l_1
  \end{array}\right\}
  \left \{ \begin{array}{ccc}
        l_1  & l_3 & k_2 \\
        l_4 & l_2  & k_1
  \end{array}\right\}
  \left \{ \begin{array}{ccc}
        k_2  & j_s & j_q \\
        j_b & l_2  & l_4
  \end{array}\right\}
                                      \nonumber \\
&& \times X_{k_2}(pqrs) \bm{\tau}^{rs}_{ab}(l_3,l_4)
  + \sum_{cdk_2} \sum_{l_3 l_4}(2 l_1 + 1)(2 l_2 + 1) 
    (-1)^{j_a + j_p + j_b + j_q + k_1 + k_2 + l_3 + l_1}
     \left \{ \begin{array}{ccc}
        k_2  & j_a & j_c \\
        j_p & l_3  & l_1
  \end{array}\right\}
                                      \nonumber \\
 &&\times \left \{ \begin{array}{ccc}
        l_2  & l_1 & k_1 \\
        l_3 & l_4  & k_2
  \end{array}\right\}
  \left \{ \begin{array}{ccc}
        k_2  & j_d & j_b \\
        j_q & l_2  & l_4
  \end{array}\right\}                   
  X_{k_2}(cdab) \bm{\tau}^{pq}_{cd}(l_3,l_4).
\end{eqnarray}

\end{widetext}


\end{document}